\documentclass[a4paper,openany,11pt]{article}
\usepackage{}
\usepackage{amssymb}
\usepackage{amsfonts}
\usepackage{stmaryrd}
\usepackage{pifont}
\usepackage{amsmath}
\usepackage{multirow}
\usepackage{natbib}
\setcitestyle{authoryear,open={(},close={)}}
\usepackage{amsfonts,amssymb,amsmath}
\usepackage{subfigure}
\usepackage[dvips]{graphics,color}
\usepackage{mathrsfs}
\def \bec{\begin{center}}
\def \enc {\end{center}}
\def \bee {\begin{eqnarray*}}
\def \ene {\end{eqnarray*}}
\def \beq {\begin{equation}}
\def \enq {\end{equation}}
  
\def\no{\noindent}

\newtheorem{theorem}{Theorem}
\newtheorem{remark}{Remark}
\newtheorem{example}{Example}

\begin{document}

\renewcommand{\theequation}%
{\arabic{equation}}

\renewcommand{\thetheorem}{\arabic{theorem}}
\renewcommand{\theremark}{\arabic{remark}}
\renewcommand{\theexample}{\arabic{section}.\arabic{example}}

{ \centering \no{{\bf \Large
Partial Penalized Likelihood Ratio Test under Sparse Case}\\
}}

\vskip 7mm  {\centering { Shanshan Wang$^{a}$, Hengjian Cui$^{b,*}$}\\
$^a$School of Mathematical Sciences, School of Physical and Mathematical Sciences, Nanyang Technological University, Singapore {\rm 637371}, Singapore\\
$^b$School of Mathematical Sciences $\&$ BCMIIS, Capital Normal University, Beijing, 100048, China\\
}
\let\thefootnote\relax\footnotetext{$^*$Corresponding author.}
\let\thefootnote\relax\footnotetext{E-mail addresses:
hjcui@bnu.edu.cn (Hengjian Cui).}

\vskip 9mm {\centering \large \bf Abstract\\} \vspace*{0.1in}

This work is concern with testing the low-dimensional parameters of interest with divergent dimensional data and variable selection for the rest under the sparse case.
A consistent test via the partial penalized likelihood approach, called the partial penalized likelihood ratio test statistic is derived, and its asymptotic distributions under the null hypothesis and the local alternatives of order $n^{-1/2}$ are obtained under some regularity conditions. Meanwhile, the oracle property of the partial penalized likelihood estimator also holds. The proposed partial penalized likelihood ratio test statistic outperforms the full penalized likelihood ratio test statistic in term of size and power, and performs as well as the classical likelihood ratio test statistic. Moreover, the proposed method obtains the variable selection results as well as the p-values of testing. Numerical simulations and an analysis of Prostate Cancer data confirm our theoretical findings and demonstrate the promising performance of the proposed partial penalized likelihood in hypothesis testing and variable selection.\vspace{-3mm}

\vskip 3mm

\no{\bf Key Words:} Chi-squared distribution, Hypothesis testing,
Likelihood ratio, Partial penalized likelihood, SCAD

\no{\bf MSC (2010):} 62F03, 62F05

\section{Introduction}\label{sec1}

Over the past few years there has been a great deal of attention on the problem of estimating a sparse parameter $\beta\in {\bf R}^p$ associated with the collected data  $V_1, \ldots, V_n$ being independent and identically distributed (iid)
random variables with the probability density function (pdf) $f(V, \beta)$.
There has been a considerable amount of recent work dedicated to the estimation problem under the sparsity scenario, both in terms of computation and theory. A comprehensive summary of the literature in either category would be too long for our purposes here, so we instead give a short summary: for computational work, some relevant contributions are \cite{FanLi2001}, \cite{friedman2007pathwise,friedman2010regularization}, \cite{wu2008coordinate}, \cite{ZouLi2008}, \cite{WuLiu2009}, \cite{breheny2011coordinate},
\cite{FanLv-106}, \cite{mazumder2011sparsenet} and so on;
and for theoretical work see, e.g., \cite{tibshirani1996regression, tibshirani2011regression}, \cite{FanLi2001}, \cite{FanPeng2004}, \cite{HastieTibshirani-114}, \cite{FanLv2008}, \cite{buhlmann2011statistics}. Generally speaking, with a few exceptions, existing theories only handle the
problem of variable selection and estimation simultaneously,
however, few of them addresses the problem of assigning uncertainties, statistical significance or confidence. As pointed out in \cite{LockhartTaylor-83}, there are still major gaps in our understanding of these regularization methods as an estimation procedure, and in many real applications, a practitioner will undoubted seek some sort of inferential guarantees for his or her variable selection procedures-but, generically, the usual constructs like p-values, confidence intervals, etc., do not exist for these estimates, especially for the zero coefficients excluded by some variable selection procedures.
In this sense, developing statistical inference methods under the sparse case is necessary.

More recently, there is a growing literature dedicated to statistical inference in the high-dimensional settings, and important progress has certainly been achieved. See \cite{WassermanRoeder-92} and \cite{MeinshausenMeier-86} for variable selection and p-value estimation based on sample splitting; stability selection in \cite{meinshausen2010stability} and \cite{ShahSamworth-94}; p-value for parameter components in lasso and ridge regression in \cite{ZhangZhang-91} and \cite{buhlmann2013statistical}; optimal confidence regions and tests for single or low-dimensional components in a high-dimensional model in \cite{van2013asymptotically} and \cite{JavanmardMontanari-85,JavanmardMontanari-90}; perturbation resampling-based procedures in \cite{MinnierTian-93}; conservative statistical inference after model selection and classification in \cite{berk2010statistical} and \cite{laber2011adaptive}, respectively;  the covariance test for Lasso model in \cite{LockhartTaylor-83}; and references therein. Apart form the aforementioned literature, our investigation is largely motivated by some doubts pertaining to variable
selection procedure. For example, if one concerns how a given genes expressions (Usually, these gens expressions of interested is known
in advance due to some prior knowledge or else) among a great amount of genes expressions affect the survival times of patients, while the variable selection procedure excludes these variables, then one need to answer the question: with how much
probability that these genes expressions have no influence on the survival times of patients (or with how much probability that the
variable selection procedure excluded these variables). Moreover, if the gens expressions of interest are included in the variable selection procedure, one hope to verify it, and to obtain a test procedure that is consistent with variable selection results.
Finally, the proposed method can also perform variable selection for the remaining gens expressions,
since the rest usually occupy the majority and also satisfy sparsity assumption. Thus, the focus of current paper is to propose a method to achieve these multiple objectives simultaneously, performing a consistent hypothesis testing for the variables of interest and variable selection for the remaining sparse ones.

Specifically, consider a canonical instance of a inference problem under the sparse case,
namely that performing hypothesis testing for a sub-vector of parameter $\beta_0\in {\bf R}^p$
based on iid observations $V_1, \ldots, V_n$ with pdf $f(V,\beta_0)$.
The null hypothesis of interest is formulated as
\begin{equation}\label{hypothesis}
H_0: \beta_{01}=0\ \ \ \ \ \mbox{vs.}\ \ \ \ \ H_1:
\beta_{01}\neq 0.
\end{equation}
Here
the true sparse parameter vector
$\beta_0=(\beta^{T}_{01},\beta^{T}_{02})^{T}$, where
$\beta_{01}\in {\bf R}^d$ is the parameter of interest with fixed and known $d\ll p$,
and its complement $\beta_{02}\in{\bf R}^{p-d}$ is sparse.
Without loss of generality, let the sparse parameter $\beta_{02}=(\beta^T_{021},\beta^T_{022})^{T}$ with the first $s$
components of $\beta_{02}$, denoted by $\beta_{021}$, do not vanish
and the remaining $p-d-s$ coefficients, denoted by $\beta_{022}$, are 0.
Rewritten $\beta_0=(\beta^{\mathcal{D}T}_0,\beta^{\mathcal{I}T}_0)^{T}$, where we refer to
$\beta^{\mathcal{D}}_{0}=(\beta^{T}_{01},\beta^{T}_{021})^{T}\in {\bf R}^{d+s}$ as an active parameter vector and its complement $\beta^{\mathcal{I}}_0=\beta_{022}=0\in{\bf R}^{p-d-s}$ as an inactive parameter vector.

For hypothesis problem (\ref{hypothesis}), the classical likelihood ratio (OLR) test proposed by \cite{neyman1928useI,neyman1928useII} is a primary one, and has been proved to have desirable properties in the literature. For example, \cite{wilks1938large} showed that the OLR test has a limiting central chi-square distribution with
$d$ degrees of freedom ($\chi^2_d$) under $H_0$, and subsequently \cite{wald1943tests} proved that the OLR test converges uniformly in distribution to the noncentral chi-square distribution under the local alternatives of order $n^{-1/2}$, i.e., $H_1: \beta_{01}=\theta+\delta n^{-1/2}$ with $\delta$ is a known $d\times 1$ vector, facilitating the power calculation. However, for the sparse parameter $\beta_{02}$, none of the estimated parameters is exactly zero in the estimation scheme of the classical likelihood (OL) method, leaving all covariates in the final model.
Consequently the OL method is incapable of selecting important variables, leading to bad predictability and estimation accuracy. And this drawback becomes worse as the sparsity level increases. To achieve variable selection for $\beta_{02}$, \cite{FanLi2001} studied the oracle properties
of nonconcave penalized likelihood estimators in the finite-dimensional setting. Their results were extended later by \cite{FanPeng2004} to the setting of $p = o(n^{1/5})$ or $o(n^{1/3})$. Yet, their penalized likelihood (PL) method may not distinguish nonzero component when it is near zero, i.e., $\delta n^{-1/2}$ for fix $\delta\neq 0$, since their proposed method is based on the condition that the nonzero component deviates from zero
at a greater rate than $O(n^{-1/2})$ for $p$ fixed and $O((n/p)^{-1/2})$ for $p$ divergent, respectively.
Moreover, \cite{FanPeng2004} also proposed the penalized likelihood ratio (PLR) test for the linear hypothesis testing concerning the nonzero components, and investigated its asymptotic null distribution. However, their proposed PLR test only applies to the hypothesis testing for the remarkable nonzero components, and may not perform inference for the zero components, i.e., the hypothesis (\ref{hypothesis}). More specifically, if we apply the PLR test for the hypothesis (\ref{hypothesis}), the estimate $\hat\beta_1$ will shrink to zero when the true value $\beta_{01}$ is near zero, owing to the estimation scheme of the full penalization, consequently leading to a conservative test. This will inevitably increase the type II error (accept $H_0$ under the alternative hypothesis $H_1$) of PLR test. Fortunately, imposing no penalization on $\beta_1$ will protect it against shrinking to zero, and obtain a consistent test. This motivates us to consider the partial penalization, and see the toy example in Section \ref{sec2:sub2} to gain more insights about the motivation for the partial penalization.

Thus, in this article we take a different way, namely by adopting
the partial penalization instead of the full penalization, we consider both problems of variable selection and
hypothesis testing for (\ref{hypothesis}), in the hope that the proposed method will possess the advantages of both OL and PL methods. Specifically, we propose the partial penalized likelihood (PPL) method to perform variable selection for sparse parameter $\beta_2$, establishing its oracle property; meanwhile, for the hypothesis (\ref{hypothesis}), we derive a consistent test, called the partial penalized likelihood ratio (PPLR) test, and under some regularity conditions we establish that the PPLR test converges in distribution to $\chi^2_d$ under $H_0$ (Theorem \ref{theorem1}) and $\chi^2_d(\gamma)$ with the noncentral parameter $\gamma$ depending on $\delta$ under the local alternatives of order $n^{-1/2}$ (Theorem \ref{theorem2}), respectively. In this sense, our proposed a consistent test performs as well as the OLR, and the PPL method is also capable of selecting important variables as PL method, achieving better predictability and estimation accuracy. Overall, the main contribution of this paper is to propose the idea of partial penalization as well as a consistent test for (\ref{hypothesis}), demonstrating its promising advantage in variable selection and hypothesis testing.

The rest of the paper is organized as follows. In Section \ref{sec2}, we first
briefly review the penalized likelihood method and the penalized
likelihood ratio test statistic proposed in \cite{FanPeng2004}, and then illustrate our motivation via a toy example. For hypothesis (\ref{hypothesis}),
we propose the partial penalized likelihood ratio test statistics in the framework
with $p$ diverging with $n$ in Section \ref{sec3:sub1}, together with its asymptotic properties. In Section \ref{sec3:sub2}, we describe the algorithm and discuss selection of tuning parameters. Numerical comparisons
and simulation studies are conducted in Section \ref{sec4}. An application to
the Prostate Cancer data is given in Section \ref{sec5}. Some discussion is
given in Section 6. Technical proofs are relegated to the Appendix.

\section{Full Penalized likelihood and a toy example}
\label{sec2}

In Section \ref{sec2:sub1}, we first briefly review existing results for nonconcave penalized likelihood approach,
and more details can be found in the work of \cite{FanLi2001} and \cite{FanPeng2004}. The more familiar reader may skip Section \ref{sec2:sub1}. Then in Section \ref{sec2:sub2}, we simply show a toy example for the possible problems existing in the full penalized likelihood method, as well as the better illustration of the idea of the partial penalization.

\subsection{Full penalized likelihood and its tests}\label{sec2:sub1}

Recall that $\log f(V, \beta)$ is the underlying
likelihood for random vector $V$, and $V_1, \cdots, V_n$, are iid samples with pdf $f(V, \beta_0)$.
Let $L_n(\beta)=\sum_{i=1}^n\log f(V_i,\beta)$ be the log-likelihood
function, and let $p_{\lambda}(|\beta_j|)$ be a nonconcave penalized function with a
tuning parameter $\lambda\geq 0$. As discussed in \cite{FanLi2001}, the penalized
likelihood estimator $\hat\beta$ then maximizes the penalized likelihood
\begin{equation}\label{eq1}
   Q_n(\beta|V)=L_n(\beta)-n\sum_{j=1}^pp_{\lambda}(|\beta_j|).
\end{equation}

For penalty function $p_{\lambda}(\cdot)$, many variable
selection-capable penalty functions have been proposed. A well known
example is the Lasso penalty \citep{tibshirani1996regression, tibshirani2011regression}. Among many
others are the SCAD penalty \citep{FanLi2001}, elastic-net penalty
\citep{zou2005regularization}, adaptive $L_1$ \citep{Zou2006}, and minimax
concave penalty \citep{Zhang-101}. In particular, \cite{FanLi2001}
studied the choice of penalty functions in depth. They proposed a
unified approach via nonconcave penalized likelihood to
automatically select important variables and simultaneously estimate
the coefficients of covariates. In this paper, we will use the SCAD
penalty for our method whenever necessary, although other penalties
can also be used. Specifically, the first derivative of SCAD
penalty satisfies
$$p'_{\lambda}(|\beta|)=\lambda\ \mbox{sgn}(\beta)\{I(|\beta|\leq
\lambda)+\frac{(a\lambda-|\beta|)_{+}}{(a-1)\lambda}I(|\beta|>\lambda)\},
\ \mbox{for some}\ a>2,$$ and $(s)_+=s$ for $s>0$ and 0 otherwise.
Following Fan and Li \cite{FanLi2001}, we set $a=3.7$ in our work.
The SCAD penalty is non-convex, leading to nonconvex
optimization. For the non-convex SCAD penalized optimization, \cite{FanLi2001} proposed the local quadratic approximation; \cite{ZouLi2008} proposed the local linear approximation; \cite{WuLiu2009}
presented the difference convex algorithm; \cite{breheny2011coordinate}
investigated the application of coordinate descent algorithms to
SCAD and MCP regression models. In this work, whenever necessary we
use the idea of coordinate descent algorithm to solve the SCAD
penalized optimization.

With a slight abuse of notation, only in this section let
$\beta_{0}=(\beta^T_{01},\beta^T_{02})^{T}$ with the first $s$
components of $\beta_{0}$, denoted by $\beta_{01}$, do not vanish
and the remaining $p-s$ coefficients, denoted by $\beta_{02}$, are 0.
For the nonconcave penalized likelihood estimator $\hat\beta$, under some
regularity conditions, \cite{FanLi2001} and \cite{FanPeng2004}
established its oracle properties in the framework with dimension $p$ fixed
and $p$ divergent, respectively. And \cite{FanPeng2004} also investigated the linear hypothesis $H_0: A\beta_{01}=0\
\mbox{versus}\  H_1: A\beta_{01}\neq 0$, where $A$ is a $q\times s$
matrix and $AA^{T}=I_q$ with a fixed $q\leq s$, and formulated the penalized likelihood ratio test statistic
as $T_n=2\{\sup_{\Omega}Q_n(\beta|V)-\sup_{\Omega,
A\beta_{1}=0}Q_n(\beta|V)\}$, where denote by $\Omega$ the
parameter space for $\beta$. Under $H_0$, with some additional conditions on the
penalty function $p_{\lambda}(\cdot)$ as in \cite{FanPeng2004},
they obtained that $T_n \rightarrow \chi^2_q$ in distribution as $n\rightarrow\infty$.

For the zero components $\beta_{02}$, their oracle property only
shows that $\hat{\beta}_{2}=0$ with probability tending to 1, as
$n\rightarrow\infty$. However, someone may suspect the assertion
$\beta_{2}=0$, or want to know with how much probability that
one of components of $\beta_{2}$ equals to zero for a given sample
size, these questions actually involve the aspects of statistical
hypotheses testing. The full penalized likelihood ratio test
statistic $T_n$ only involves the linear hypotheses for nonzero
components $\beta_{01}$, and the conclusion for $T_n$ under $H_0$
may not hold for some special $A$. For example, when $A=I_s$, it
follows that $\beta_0=0$ under $H_0$, then $T_n=o_p(1)$, since
$\hat{\beta}=0$ with probability tending to 1 (oracle property). Or
$A=e_j$, where $e_j$ is a $s\times 1$ vector with the $j$th
component is 1 and 0 otherwise, then under $H_0$, $T_n$ may be not
asymptotically $\chi^2$ distributed. We will demonstrate this
phenomena in the simulation studies.

\subsection{A toy example}\label{sec2:sub2}

\begin{figure}[!h]
\begin{center}
\subfigure[]{
\label{fig0.sub1}
 \resizebox{5cm}{5cm}{\includegraphics{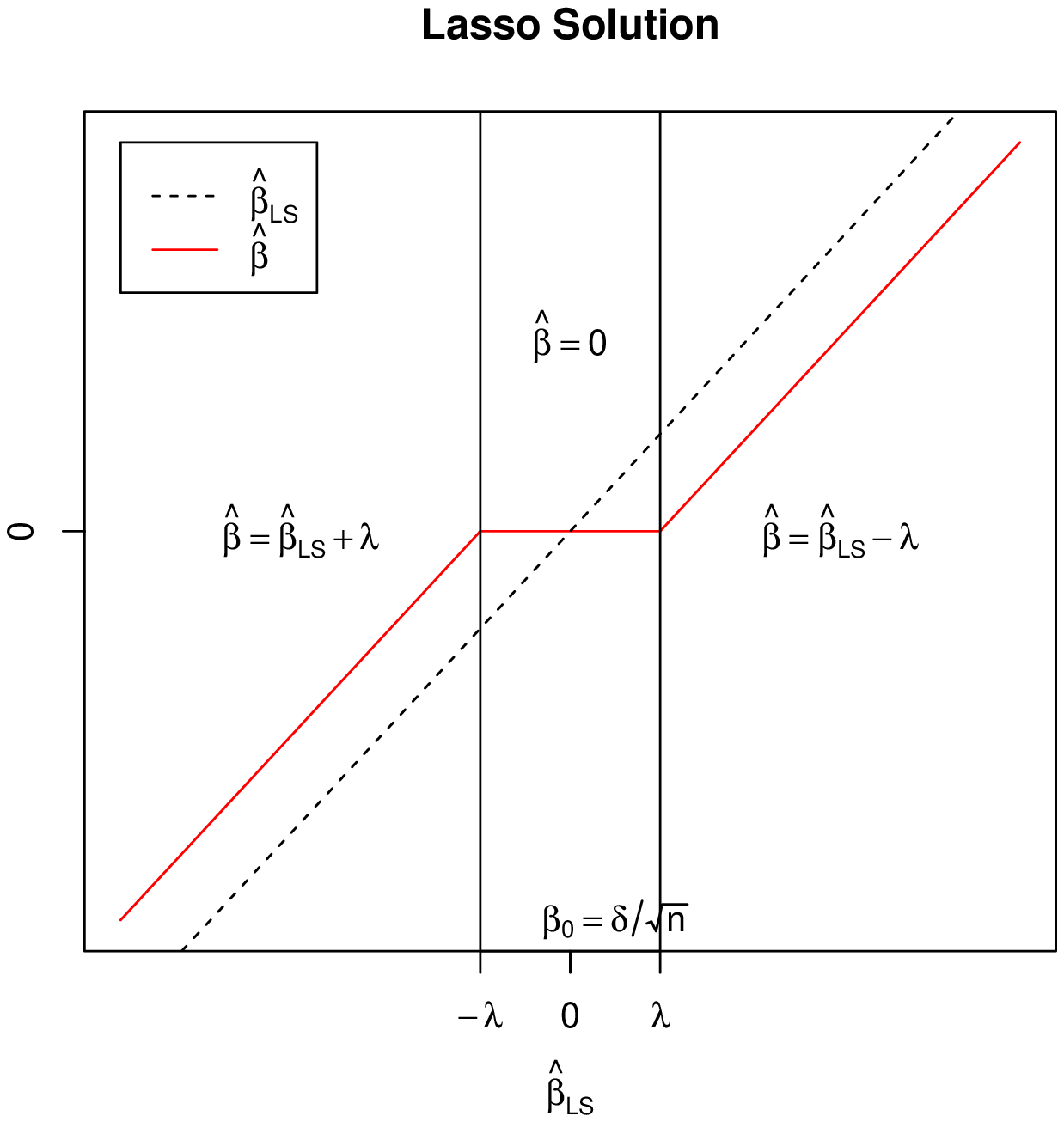}}}
 \subfigure[]{
\label{fig0.sub2}
 \resizebox{5cm}{5cm}{\includegraphics{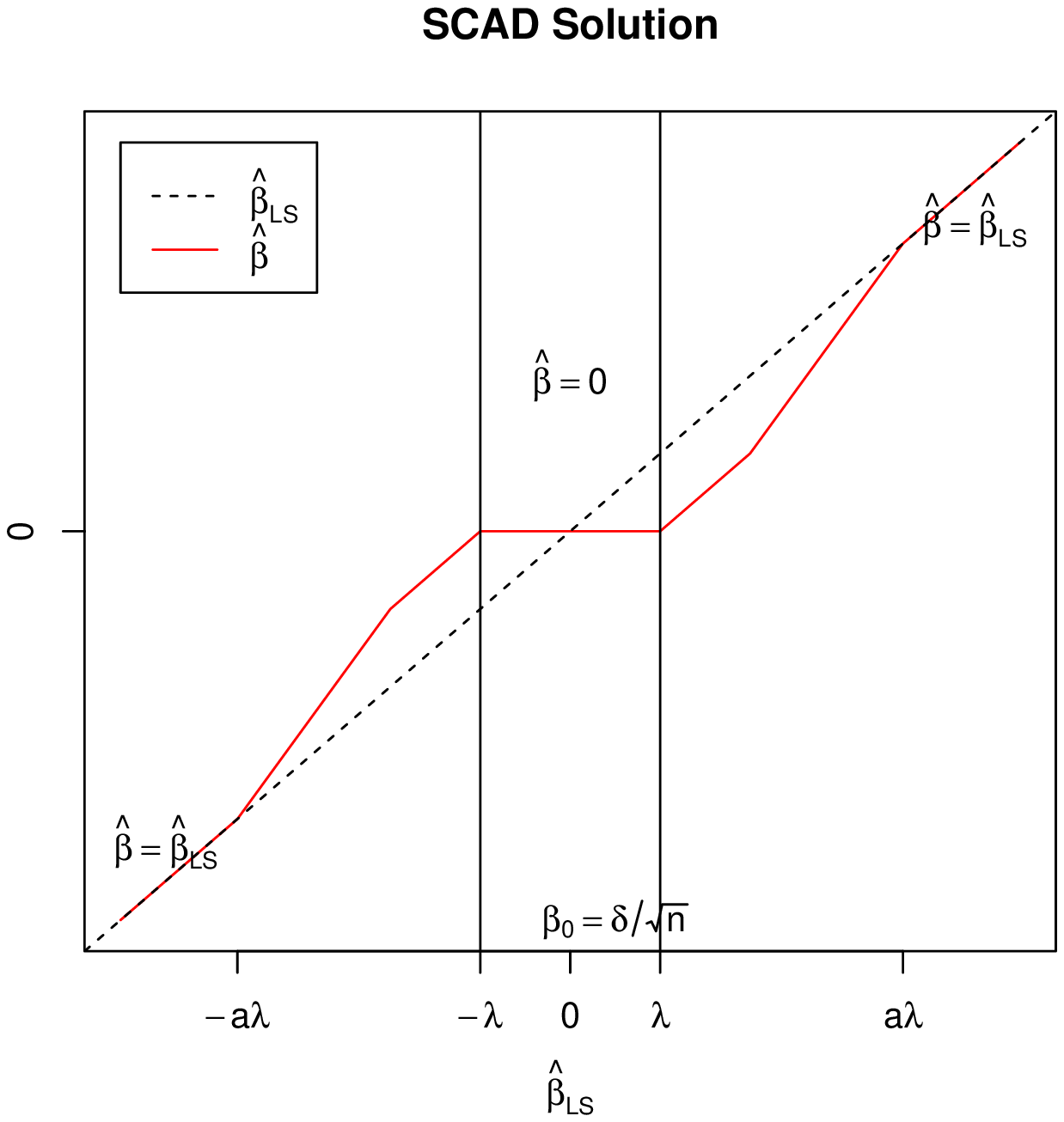}}}
 \subfigure[]{
\label{fig0.sub3}
 \resizebox{5cm}{5cm}{\includegraphics{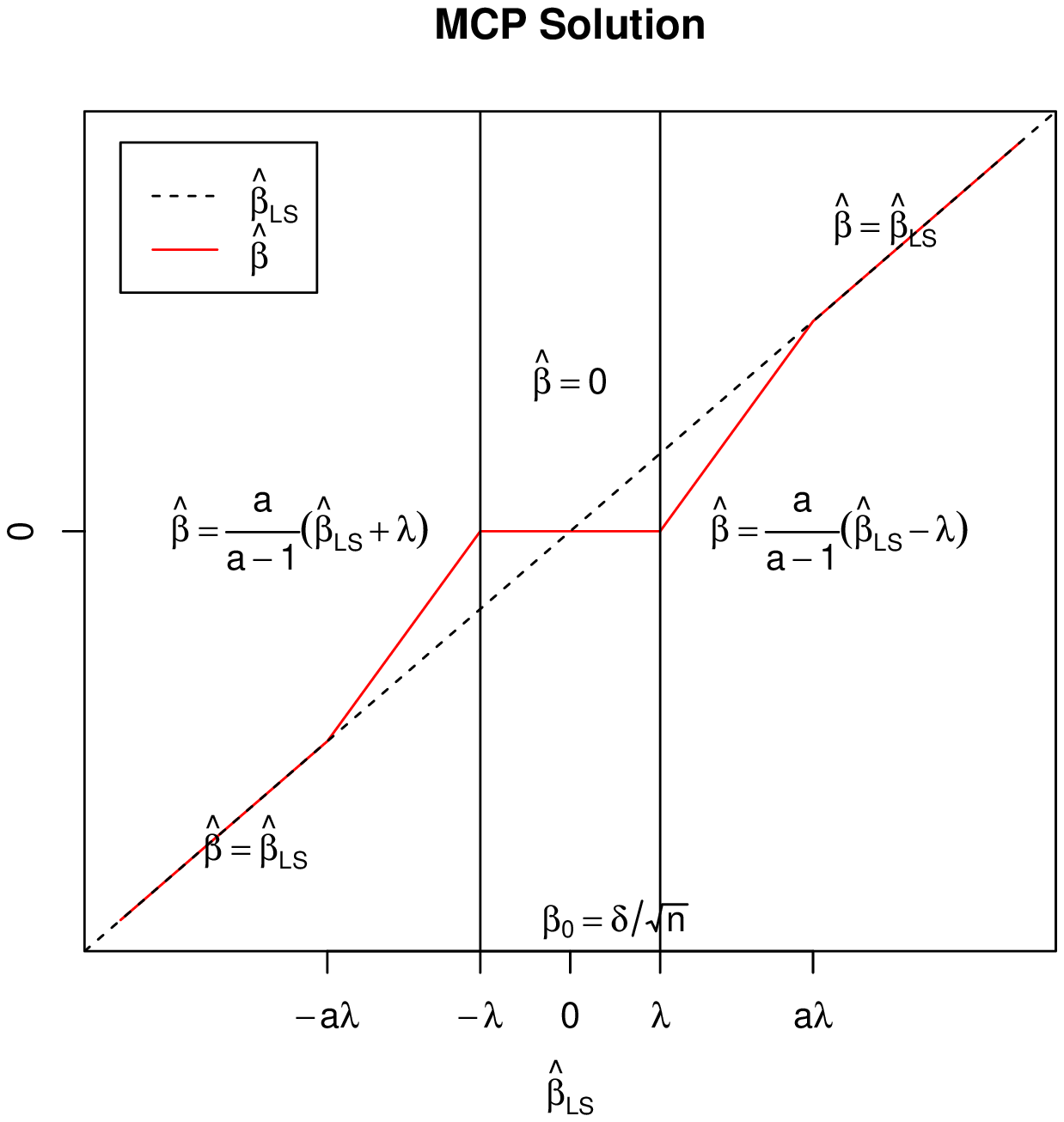}}}
\end{center}\vspace{-0.8cm}
\begin{center}
 \caption{Given $\lambda$, plots of the penalized least square estimate $\hat\beta$ versus the least square estimate $\hat\beta_{LS}$ in Eq. (\ref{betahat}) for the Lasso (a), SCAD (b) and MCP (c) penalties, respectively.}\label{fig0}
\end{center}
\end{figure}

Before we present the main approach, here we simply show a toy example for the better illustration of the partial penalization. As in \cite{FanLi2001}, consider the linear regression model $Y=X\beta+\varepsilon$, where assume that the error vector $\varepsilon=(\varepsilon_1, \cdots, \varepsilon_n)^T$ with $\varepsilon_i\overset{iid}\sim N(0,1)$,  the $n\times 1$ response vector $Y=(Y_1,\ldots,Y_n)$, and the $n\times p$ matrix $X=(X_1,\ldots,X_n)^T$ satisfy
\begin{equation}\label{standarization}
  \sum_{i=1}^nY_i=0,~~\sum_{i=1}^nX_{ij}=0,~~\sum_{i=1}^n X^2_{ij}=n.
\end{equation}
It is well known that the classical maximum likelihood estimate of $\beta$ corresponds to the least square estimator $\hat\beta_{LS}=n^{-1}X^TY$, and the maximum penalized likelihood estimate of $\beta$ defined in (\ref{eq1}) corresponds to the penalized least square estimator, denoted by $\hat\beta$, and under the assumption (\ref{standarization}), it holds that \begin{equation}\label{betahat}
                                         \hat\beta=\mbox{arg}\min_{\beta} \frac{1}{2n}\|Y-X\beta\|^2+\sum_{j=1}^pp_\lambda(|\beta_j|)                                          \propto \frac{1}{2}\|\hat\beta_{LS}-\beta\|^2+\sum_{j=1}^pp_\lambda(|\beta_j|).\end{equation}
For given $\lambda$, Figure \ref{fig0} shows the plots of the penalized least square estimator $\hat\beta$ versus the least square estimator $\hat\beta_{LS}$ in Eq. (\ref{betahat}) for the Lasso (a), SCAD (b) and MCP (c) penalties, respectively. When the true parameter $\beta_0$ is near zero, i.e., $\beta_0=\delta/\sqrt{n}$, then $\beta_0$ fall in the interval $(-\lambda,\lambda)$ with high probability tending to 1 from Figure \ref{fig0}, thus these three variable selection procedures all result in $\hat\beta=0$ owing to its penalization scheme. Consequently, when perform the hypothesis $H_0: \beta_0=0$, the estimate $\hat\beta$ will shrink to zero when the true value $\beta_{0}$ is near zero, leading to a conservative test. This will inevitably increase the type II error (accept $H_0$ under the alternative hypothesis $H_1: \beta_0=\delta/\sqrt{n}$). To obtain a consistent test, we consider the partial penalization in Section \ref{sec3}.

\section{Partial penalized likelihood and its test}
\label{sec3}

For the sparse parameter, variable selection through regularization
has proven to be effective, and possesses desirable oracle
properties under some regularized conditions. However, as discussed
in Section \ref{sec1} as well as that at the end of Section \ref{sec2}, it is
necessary to develop a new approach to deal with the hypothesis test
concerning variable selection results.
For hypothesis (\ref{hypothesis}), we derive
a consistent test procedure
in the framework with $p$ divergent in Section \ref{sec3:sub1}. And the
implementation of the test procedure and choice of tuning parameter
is described in Section \ref{sec3:sub2}. Throughout this paper, it is important to note that the quantities $p$ and $\lambda$ can depend on the sample size $n$, and we have suppressed this dependency for natational simplicity.

\subsection{Partial penalized likelihood ratio test}
\label{sec3:sub1}

Recall that $V_1, \cdots, V_n$, are iid random variables with pdf $f(V, \beta_0)$, and the parameter $\beta_0$ is the same as that in Section \ref{sec1}. For the
hypothesis problem (\ref{hypothesis}), we define a partial penalized
likelihood ratio test statistic as
\begin{equation}\label{pplr1}
    T_{n}=2\{\sup_{\Omega}PQ_n(\beta|V)-\sup_{\Omega, \beta_1=0}PQ_n(\beta|V)\},
\end{equation}
where \begin{equation}\label{eq2}
   PQ_n(\beta|V)=L_n(\beta)-n\sum_{j=d+1}^pp_{\lambda}(|\beta_j|)
\end{equation}
is the partial penalized likelihood function, with
$\hat{\beta}=\mbox{argmax}_{\beta}PQ_n(\beta)$ being the
partial penalized likelihood estimator.

\begin{remark}\quad Here in (\ref{eq2}), instead of full penalized, we propose
partial penalized with $\beta_1$ nonpenalized. This will protect $\beta_1$ against
shrinking to zero when the true value $\beta_{01}$ is
zero or near zero, and to make further statistical inference. In fact, $\beta_{01}$ is the parameter of interested and $\beta_{02}$ is sparse, with partial penalized in (\ref{eq2}), we can not only protect $\beta_1$, but also perform variable selection for $\beta_2$.
\end{remark}

\begin{remark}\quad
For the linear hypothesis $H_0: A\beta_{0}=0\
\ \mbox{vs.}\ \ H_1: A\beta_{0}\neq0$, where $A$ is a $d\times p$
matrix and $AA^{T}=I_d$ for a fixed $d\ll p$. This problem includes
the problem of testing simultaneously the significance of a few
parameters. Let $B$ be a $(p-d)\times p$ matrix which satisfies
$BB^{T}=I_{p-d}$ and $AB^{T}=0$. That is, the linear space spanned
by rows of $B$ is the orthogonal complement to the linear space
spanned by rows of $A$. Let $\tilde{\beta}=\tilde{A}\beta$ with
$\tilde{A}=(A^T,B^T)^T$ satisfying $\tilde{A}\tilde{A}^{T}=I_p$,
then the linear hypothesis $H_0: A\beta_{0}=0\ \ \mbox{vs.}\ \ H_1:
A\beta_{0}\neq0$ can be reformulated as $H_0: \tilde{\beta}_{01}=0\
\ \mbox{vs.}\ \ H_1: \tilde{\beta}_{01}\neq0$, where
$\tilde{\beta}_{01}$ is the first $d$ components of parameter
$\tilde{\beta}_0$. Then the partial penalized likelihood function in
(\ref{eq2}) can be defined as $
   PQ_n(\tilde{\beta}|V)=L_n(\tilde{\beta})-n\sum_{j=d+1}^pp_{\lambda}(|\tilde{\beta}_j|)
$ with $L_n(\tilde{\beta})=\sum_{i=1}^n \log f(V_i,
\tilde{A}^{-1}\tilde{\beta})$. And the corresponding partial
penalized likelihood ratio test statistic $T_n$ can also be
constructed.
\end{remark}

Denote $\tilde{p}_{\lambda}(\cdot)$ as a working penalty function, where
$\tilde{p}_{\lambda}(|\beta_j|)=0$ for $1\leq j\leq d$, and
$\tilde{p}_{\lambda}(|\beta_j|)=p_{\lambda}(|\beta_{j}|)$ for
$(d+1)\leq j\leq p$, then the partial penalized likelihood function
in (\ref{eq2}) can be rewritten as $Q_n(\beta|V)=L_n(\beta)-n\sum_{j=1}^p \tilde{p}_{\lambda}(|\beta_{j}|)
$, which can be seen as the penalized likelihood function with a
special penalty function $\tilde{p}_{\lambda}(\cdot)$ in
(\ref{eq1}). Therefore, it follows that the oracle property of
$\hat{\beta}$ as in Theorems 1 and 2 of \cite{FanPeng2004}'s paper
also hold. See Lemmas 1 and 2 in the Appendix.

Based on the oracle property of $\hat\beta$, we investigate the asymptotic properties of $T_n$ in (\ref{pplr1}) under $H_0$ in (\ref{hypothesis}) as well as the local alternatives $H_1:\beta_{01}=\delta n^{-1/2}$, where $\delta$ is a known $d\times 1$
vector. The following theorems drive the asymptotic null distribution and
the local alternative distribution of $T_n$, facilitating hypothesis
testing and the power calculation. It shows that the classical
likelihood theory continues to hold in the partial penalized
likelihood context.

\begin{theorem}\label{theorem1}\quad
When regularized
conditions (A)-(H), and $(E')$ and $(F')$ in the Appendix are
satisfied, under $H_0$ it holds that $T_n \rightarrow \chi^2_{d}$,
provided that $p^5/n\rightarrow 0$ as $n\rightarrow\infty$.
\end{theorem}

\begin{theorem}\label{theorem2}\quad When regularized
conditions (A)-(H), and $(E')$ and $(F')$ in the Appendix are
satisfied, if $H_1: \beta_{01}=\delta n^{-1/2}$ is true,
where $\delta$ is a $d\times 1$ vector, it holds that $T_n \rightarrow
\chi^2_{d}(\gamma)$ with the noncentral parameter $\gamma=\delta^{T}C_{11.2}\delta$, provided
that $p^5/n\rightarrow 0$ as $n\rightarrow\infty$. Where
$C_{11.2}=C_{11}-C_{12}C^{-1}_{22}C_{21}$ and $(d+s)\times
(d+s)$ matrix $I_{1}(\beta^{\mathcal{D}}_{0})=I_1(\beta^{\mathcal{D}}_{0}, 0)=\left(
                            \begin{array}{cc}
                              C_{11} & C_{12} \\
                              C_{21} & C_{22} \\
                            \end{array}
                          \right)
$, the Fisher information knowing $\beta^{\mathcal{I}}_{0}=0$, with principal submatrices $C_{11}$ and $C_{22}$ are $d\times d$
and $s\times s$, respectively.
\end{theorem}

\begin{remark}\quad The condition $p^4/n\rightarrow 0$ in
Theorem \ref{theorem1} or $p^5/n\rightarrow 0$ in Theorem \ref{theorem2} as $n\rightarrow
\infty$ seems somewhat strong, where the rate on $p$ should not be
taken as restrictive because our proposed method is studied in a
broad framework based on the log-likelihood function. Since no
particular structural information is available on the log-likelihood
function, establishing the theoretical result is very challenging,
so the strong regularity conditions are needed and the bound in the
stochastic analysis are conservative. This is also the case in \cite{FanPeng2004}. By refining the structure of the log-likelihood
function, the restriction on dimensionality $p$ can be relaxed.
Another reason is the stronger conditions on the likelihood
function, which facilitate the technical proofs, yet may bring
stringent assumption on $p$. Since our focus in this section is to
demonstrate our proposed method may be applicable in the framework
with $p$ growing with $n$. Yet, the question that how sharpest the
dimension $p$ may be growing with $n$ isn't addressed in this
paper, which we will consider in the future work. Thus, keep in mind
that the framework presented in this paper is applicable only where
the sample size is larger that the dimension of the parameter. When
that is violated, preliminary methods such as sure independence
screening \cite{FanLv2008} may be used to reduce the
dimensionality, and then adopt our proposed method.
\end{remark}

\subsection{Tuning and Implementation}
\label{sec3:sub2}

In this Section, we describe an efficient coordinate descent
algorithm for the implementation of the proposed method, and
discuss the selection of tuning parameters.

The idea of coordinate optimization for penalized
problems was proposed by \cite{fu1998penalized}, and was demonstrated by \cite{friedman2007pathwise} and \cite{wu2008coordinate} to be efficient
for large-scale sparse problems. Recently, various authors,
including \cite{FanLv2010}, \cite{breheny2011coordinate}, and \cite{mazumder2011sparsenet}
generalized this idea to regularized
regression with various penalties and showed that it was an
attractive alternative to earlier proposals such as the local
quadratic approximation \citep{FanLi2001} and the local linear
approximation \citep{ZouLi2008}.

To maximize objective function $PQ_n(\beta|V)$ in (\ref{eq2}), the
coordinate descent method maximizes the objective function in one
coordinate at a time and cycles through all coordinates until
convergence. For fixed $\lambda$, cyclically for $j=1, \ldots, p$,
update the $j$th component $\hat\beta_j(\lambda)$ of
$\hat\beta(\lambda)$ by the univariate maximizer of
$PQ_n(\hat\beta(\lambda)|V)$ with respect to $\hat\beta_j(\lambda)$
until convergence. Then this produces a solution path
$\hat\beta(\lambda)$ over a grid of points $\lambda$, then the optimal regularization
parameter $\lambda$ can be chosen by  minimizing the following \mbox{BIC} type criteria
motivated by \cite{WangLi2009},
\begin{equation}\label{tune}
    \mbox{BIC}(\lambda)=-2PQ_n(\hat\beta(\lambda)|V)+C_n\log(n)\mbox{df}_{\lambda},
\end{equation}
where $\hat\beta(\lambda)$ is the partial penalized likelihood
estimate of $\beta$ with regularization parameter $\lambda$;
$\mbox{df}_{\lambda}$ is the number of nonzero coefficients in
$\hat\beta(\lambda)$; $C_n$ is a scaling factor diverging to
infinity at a slow rate \cite{WangLi2009} for $p\rightarrow
\infty$, and they suggested that
$C_n=\max\{\log\log p,1\}$ seemed to be a good choice. However, a
rigorous proof of the consistency of this \mbox{BIC} for partial
penalized likelihood merits further investigation. Fortunately, the
BIC type criterion defined in (\ref{tune}) usually selects the
tuning parameter satisfactorily and identifies the true model
consistently in our simulation studies.

Finally, it is important to keep in mind that the optimal
regularization parameter $\lambda$ should be the same for maximizing
$PQ_n(\beta|V)$ within the full parameter space and the subspace
specified by the null hypothesis in (\ref{hypothesis}), when we
calculate test statistics $T_n$ in (\ref{pplr1}).
In fact, we adopt the aforementioned \mbox{BIC} to choose the
optimal regularization parameter $\lambda$ when maximize
$PQ_n(\beta|V)$ within the full parameter space, and then for the
chosen $\lambda$ we maximize $PQ_n(\beta|V)$ within the subspace
specified by the null hypothesis in (\ref{hypothesis}).

\section{Numerical comparisons}\label{sec4}

We present simulation results to illustrate the usefulness of the
partial penalized likelihood ratio (PPLR) test, and to compare the
finite-sample performance with the penalized likelihood ratio (PLR) test
and the classical likelihood ratio (LR) test in terms of model
selection accuracy and power. That is, we first assess the
performance of the partial penalized likelihood (PPL), the penalized
likelihood (PL) and the ordinary likelihood (OL) in terms of
estimation accuracy and model selection consistency. Then we
evaluate empirical size and power of these three test methods. Here
we set $d=1$, and the BIC type criterion defined in (\ref{tune}) is
used to estimate the optimal tuning parameter $\lambda$ in the
smoothly clipped absolute deviation (SCAD).
And we simulate 1000 samples of size $n=100, 200, 400$
and 800 with $p=11,20,30$ and $41$ from the following two examples:

\begin{example}\label{example1}\quad (Linear Regression)
$Y=X^{T}\beta+\sigma\varepsilon$, where we set $\sigma=1$,
$\varepsilon$ follows a standard normal distribution, $X\sim N(0, I_{p})$, and the true value
$\beta_{0}=(\beta_{01},3,1.5,2,1,0,\cdots,0)^{T}\in
{\bf R}^{p}$ with $\beta_{01}$ is the parameter of interest, and will
be specified as different true values whenever necessary in the
following simulations. All covariates are standardized. We consider
the null hypothesis $H_0: \beta_{01}=0$ and the local
alternatives $H_1: \beta_{01}=\delta n^{-1/2}$.
\end{example}

\begin{example}\label{example2}\quad (Logistic Regression) $Y\sim\
\mbox{Bernoulli}\{p(X^{T}\beta)\}$, where
$p(u)=\exp(u)/(1+\exp(u))$, and the covariates $X$ and $\beta$ are
the same as those in Example \ref{example1}. All covariates are standardized.
\end{example}

\begin{table}[!h]
\begin{center}
\scriptsize \caption{Results for three methods PPL, PL and OL in
Example \ref{example1} under the true values
$\beta_{0}=(\beta_{01},3,1.5,2,1,0,\cdots,0)^T$, where the first
component $\beta_{01}=\delta n^{-1/2}$ with different $\delta$.
Values shown are means (standard deviations) of each performance
measure over 1000 replicates.}\label{tab1}\vspace{-8mm}
\end{center}
\setlength\tabcolsep{1pt}
\renewcommand{\arraystretch}{1.2}
\begin{center}\scriptsize
\begin{tabular}{ccccccccccccc}
\hline
\multicolumn{2}{c}{Method}&\multicolumn{4}{c}{PPL}&\multicolumn{4}{c}{PL}&\multicolumn{2}{c}{OL}\\
\cline{1-12}
$(n,p)$&$\delta$&\multicolumn{1}{c}{$L_2$-loss}&\multicolumn{1}{c}{$L_1$-loss}&\multicolumn{1}{c}{C}&\multicolumn{1}{c}{IC}&\multicolumn{1}{c}{$L_2$-loss}&\multicolumn{1}{c}{$L_1$-loss}&\multicolumn{1}{c}{C}&\multicolumn{1}{c}{IC}&\multicolumn{1}{c}{$L_2$-loss}&\multicolumn{1}{c}{$L_1$-loss}\\
\hline
(100,11)&0.0&$0.226(0.074)$&$0.444(0.153)$&$5.317(0.780)$&$0(0)$&$0.201(0.071)$&$0.364(0.136)$&$6.181(0.849)$&$0.000(0.000)$&$0.342(0.076)$&$0.924(0.215)$\\
&1.0&$0.225(0.073)$&$0.442(0.153)$&$5.341(0.781)$&$0(0)$&$0.222(0.067)$&$0.448(0.141)$&$5.325(0.801)$&$0.727(0.446)$&$0.342(0.078)$&$0.929(0.224)$\\
&2.0&$0.228(0.072)$&$0.447(0.151)$&$5.326(0.771)$&$0(0)$&$0.257(0.069)$&$0.508(0.155)$&$5.306(0.778)$&$0.333(0.472)$&$0.345(0.080)$&$0.936(0.229)$\\
&3.0&$0.222(0.071)$&$0.438(0.149)$&$5.292(0.803)$&$0(0)$&$0.267(0.079)$&$0.518(0.164)$&$5.262(0.807)$&$0.088(0.283)$&$0.339(0.076)$&$0.921(0.220)$\\
&4.0&$0.220(0.071)$&$0.433(0.152)$&$5.344(0.747)$&$0(0)$&$0.261(0.087)$&$0.504(0.176)$&$5.326(0.768)$&$0.008(0.089)$&$0.336(0.078)$&$0.912(0.224)$\\
(200,20)&0.0&$0.160(0.048)$&$0.330(0.106)$&$14.525(1.152)$&$0(0)$&$0.145(0.048)$&$0.277(0.097)$&$15.414(1.216)$&$0.000(0.000)$&$0.340(0.054)$&$1.258(0.213)$\\
&1.0&$0.162(0.050)$&$0.331(0.110)$&$14.610(1.145)$&$0(0)$&$0.160(0.046)$&$0.336(0.103)$&$14.609(1.142)$&$0.747(0.435)$&$0.337(0.055)$&$1.251(0.215)$\\
&2.0&$0.159(0.050)$&$0.326(0.109)$&$14.552(1.159)$&$0(0)$&$0.183(0.046)$&$0.377(0.112)$&$14.536(1.173)$&$0.375(0.484)$&$0.337(0.058)$&$1.251(0.230)$\\
&3.0&$0.159(0.050)$&$0.325(0.107)$&$14.603(1.113)$&$0(0)$&$0.196(0.054)$&$0.391(0.118)$&$14.556(1.127)$&$0.107(0.309)$&$0.336(0.054)$&$1.244(0.214)$\\
&4.0&$0.159(0.050)$&$0.325(0.109)$&$14.596(1.131)$&$0(0)$&$0.192(0.062)$&$0.382(0.129)$&$14.566(1.150)$&$0.011(0.104)$&$0.337(0.055)$&$1.247(0.214)$\\
(400,30)&0.0&$0.111(0.034)$&$0.233(0.077)$&$24.157(1.292)$&$0(0)$&$0.100(0.034)$&$0.196(0.072)$&$25.095(1.307)$&$0.000(0.000)$&$0.286(0.038)$&$1.281(0.181)$\\
&1.0&$0.112(0.034)$&$0.234(0.076)$&$24.200(1.317)$&$0(0)$&$0.112(0.030)$&$0.241(0.070)$&$24.183(1.330)$&$0.805(0.396)$&$0.285(0.039)$&$1.277(0.185)$\\
&2.0&$0.114(0.033)$&$0.239(0.071)$&$24.139(1.297)$&$0(0)$&$0.132(0.031)$&$0.276(0.074)$&$24.127(1.331)$&$0.435(0.496)$&$0.287(0.038)$&$1.284(0.184)$\\
&3.0&$0.113(0.033)$&$0.237(0.075)$&$24.162(1.327)$&$0(0)$&$0.142(0.037)$&$0.289(0.083)$&$24.123(1.366)$&$0.123(0.329)$&$0.286(0.037)$&$1.281(0.176)$\\
&4.0&$0.114(0.035)$&$0.240(0.077)$&$24.107(1.326)$&$0(0)$&$0.141(0.042)$&$0.286(0.087)$&$24.074(1.337)$&$0.009(0.094)$&$0.289(0.038)$&$1.292(0.179)$\\
(800,41)&0.0&$0.080(0.025)$&$0.169(0.056)$&$34.908(1.369)$&$0(0)$&$0.072(0.024)$&$0.143(0.051)$&$35.850(1.388)$&$0.000(0.000)$&$0.234(0.026)$&$1.221(0.147)$\\
&1.0&$0.082(0.024)$&$0.174(0.054)$&$34.886(1.431)$&$0(0)$&$0.081(0.021)$&$0.177(0.050)$&$34.882(1.433)$&$0.809(0.393)$&$0.233(0.026)$&$1.213(0.146)$\\
&2.0&$0.080(0.024)$&$0.169(0.054)$&$34.919(1.361)$&$0(0)$&$0.094(0.021)$&$0.199(0.053)$&$34.910(1.383)$&$0.450(0.498)$&$0.233(0.025)$&$1.212(0.142)$\\
&3.0&$0.080(0.024)$&$0.170(0.056)$&$34.882(1.466)$&$0(0)$&$0.102(0.026)$&$0.208(0.060)$&$34.857(1.486)$&$0.136(0.343)$&$0.235(0.027)$&$1.219(0.150)$\\
&4.0&$0.081(0.024)$&$0.171(0.054)$&$34.947(1.359)$&$0(0)$&$0.102(0.030)$&$0.207(0.063)$&$34.944(1.365)$&$0.011(0.104)$&$0.233(0.026)$&$1.215(0.147)$\\
\hline
\end{tabular}
\end{center}
\end{table}

For Example \ref{example1}, first, we evaluate the performance of the resulting estimators for
the three methods in term of four measures under the different true
values $\beta_0=(\beta_{01},3,1.5,2,1,0,\cdots,0)^T$ with
$\beta_{01}=\delta n^{-1/2}$,
$\delta=0,0.5,\cdots,3.5$ and $4$, respectively. For estimation accuracy, we report the
$L_2$-loss $\|\hat\beta-\beta_0\|_2=\{(\hat\beta-\beta_0)^T(\hat\beta-\hat\beta_0)\}^{1/2}$
and $L_1$-loss $\|\hat\beta-\beta_0\|_1=\sum_{j=1}^p|\hat\beta_j-\beta_{0j}|$. The
other two measures pertain to model selection consistency:
$\mbox{C}$ and $\mbox{IC}$ refer to the number of correctly selected
zero coefficients and the number of incorrectly excluded variables, respectively. Due to space limitations, Table \ref{tab1} only summarizes the means and standard deviations of each measure over 1000 replicates for $\delta=0,1,2,3$ and $4$.
(Since none of the estimated regression coefficients is
exactly zero for the OL method, it reports no
model selection results, and we only show its estimation accuracy results in Table 1.)

From Table \ref{tab1}, we can show that the PPL method
outperforms the PL method when $\delta\neq 0$, since the average
number of incorrectly estimated zero coefficients is always greater
than 0 for the PL method. That is, the PL method may not identify
the nonzero component $\beta_{01}=\delta n^{-1/2}$ with fixed
$\delta\neq 0$; while the proposed PPL method still works. In this
way, we conjecture that the PL method can
not distinguish the nonzero component of order $n^{-1/2}$, and show
that the PPL method outperforms the PL method especially when some
of nonzero component is near zero in term of model selection. For
the estimation accuracy, the PPL method performs best among the
three methods. Therefore, if we know some components is near zero
and the rest are sparse in advance, our proposed method performs
best among the three methods, with a performance very close to that
of the oracle estimator and better estimation performance.

Next, to verify performance of the PPLR test in Theorems \ref{theorem1} and \ref{theorem2}, consider the null hypothesis $H_0:
\beta_{01}=0$, and calculate power under the local alternatives
$H_1: \beta_{01}=\delta n^{-1/2}$ for different $\delta$ and
sample size $n$, respectively. Using a nominal level $\alpha=0.05$, we documents the empirical size and the
power results in Table \ref{tab2}. From Table \ref{tab2}, for hypothesis $H_0$, the PLR test does
not work any more, while the remaining two methods still work, which
confirms Theorems \ref{theorem1} and \ref{theorem2}. From the view of this point, we can
conjecture that the PPLR test performs as well as the LR method
and outperforms the PLR test when the null parameter is zero in
terms of size and power. All of these results demonstrates the
promising performance of the PPLR test in hypothesis testing.

\begin{table}[!h]
\begin{center}
\scriptsize \caption{Empirical percentage of rejecting $H_0:
\beta_{01}=0$ for the true values under $H_1:
\beta_{01}=\delta n^{-1/2}$ with different $\delta$ and the sample size $n$ in Example \ref{example1}. The nominal
level is 5\%.}\label{tab2}\vspace{-5mm}
\end{center}
\setlength\tabcolsep{4 pt}
\renewcommand{\arraystretch}{1}
\begin{center}\scriptsize
\begin{tabular}{ccccccccccc}
\hline
$(n,p)$&\multicolumn{1}{c}{Test}&\multicolumn{1}{c}{$\delta=0.0$}&\multicolumn{1}{c}{$\delta=0.5$}&\multicolumn{1}{c}{$\delta=1.0$}&\multicolumn{1}{c}{$\delta=1.5$}&\multicolumn{1}{c}{$\delta=2.0$}
&\multicolumn{1}{c}{$\delta=2.5$}&\multicolumn{1}{c}{$\delta=3.0$}&\multicolumn{1}{c}{$\delta=3.5$}&\multicolumn{1}{c}{$\delta=4.0$}\\
\hline
(100,11)&PPLR&$0.047$&$0.084$&$0.157$&$0.311$&$0.483$&$0.705$&$0.840$&$0.935$&$0.972$\\
&PLR&$0.000$&$0.001$&$0.001$&$0.023$&$0.055$&$0.136$&$0.291$&$0.480$&$0.688$\\
&LR&$0.042$&$0.087$&$0.157$&$0.305$&$0.460$&$0.682$&$0.812$&$0.917$&$0.965$\\
(200,20)&PPLR&$0.057$&$0.096$&$0.158$&$0.314$&$0.504$&$0.686$&$0.845$&$0.936$&$0.973$\\
&PLR&$0.001$&$0.002$&$0.003$&$0.014$&$0.049$&$0.113$&$0.248$&$0.410$&$0.599$\\
&LR&$0.060$&$0.090$&$0.146$&$0.294$&$0.458$&$0.668$&$0.808$&$0.915$&$0.958$\\
(400,30)&PPLR&$0.054$&$0.076$&$0.162$&$0.305$&$0.520$&$0.711$&$0.864$&$0.936$&$0.985$\\
&PLR&$0.000$&$0.000$&$0.002$&$0.008$&$0.055$&$0.101$&$0.209$&$0.375$&$0.577$\\
&LR&$0.054$&$0.074$&$0.150$&$0.294$&$0.481$&$0.674$&$0.837$&$0.923$&$0.970$\\
(800,41)&PPLR&$0.056$&$0.087$&$0.177$&$0.314$&$0.508$&$0.716$&$0.828$&$0.928$&$0.980$\\
&PLR&$0.000$&$0.000$&$0.000$&$0.007$&$0.034$&$0.086$&$0.218$&$0.320$&$0.554$\\
&LR&$0.054$&$0.092$&$0.181$&$0.304$&$0.512$&$0.689$&$0.817$&$0.914$&$0.974$\\
\hline
\end{tabular}
\end{center}
\end{table}

\begin{figure}{}
\begin{center}
\subfigure[]{
\label{fig1.sub1}
\resizebox{5cm}{4.4cm}{\includegraphics{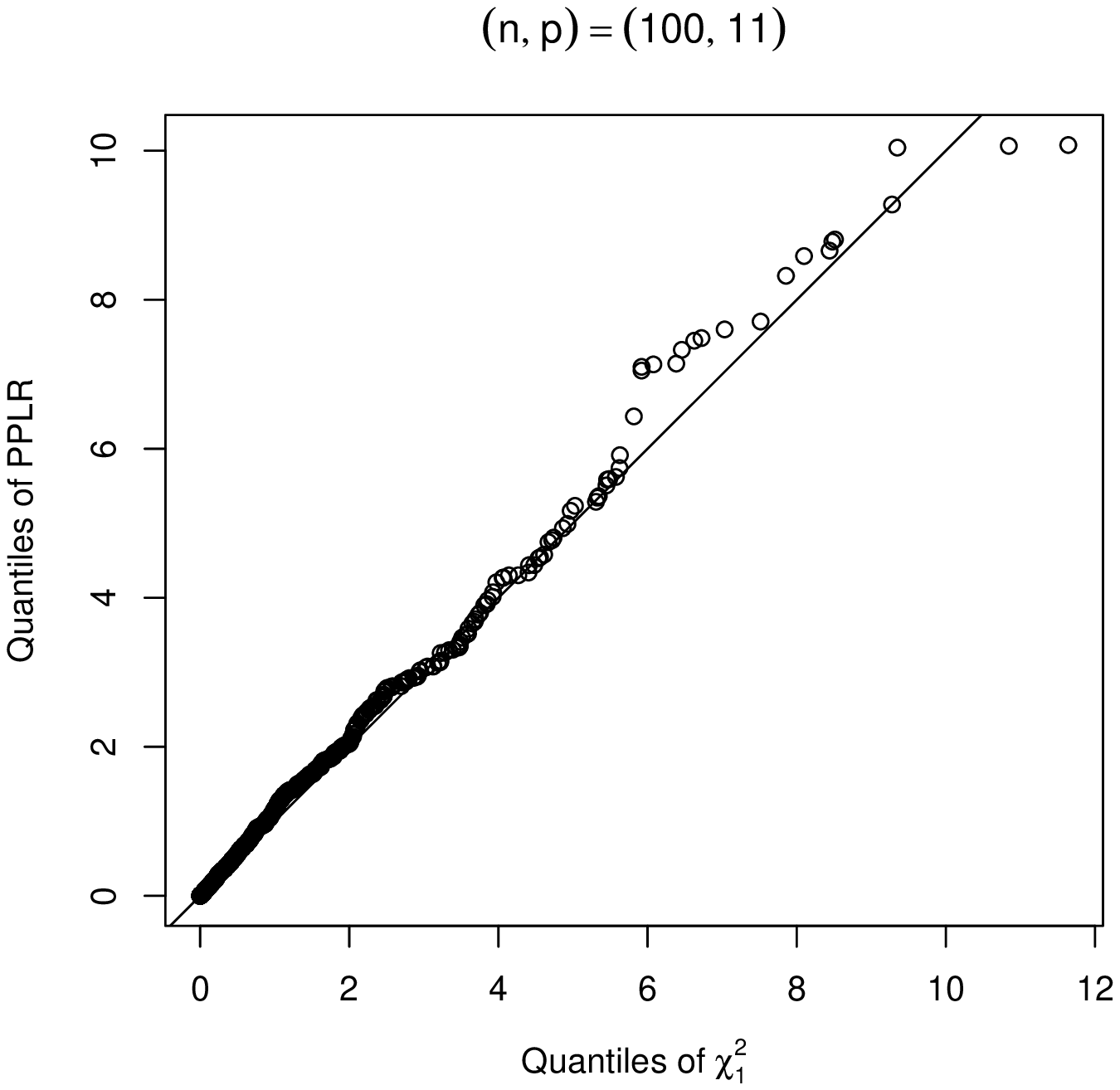}}}
\subfigure[]{
\label{fig1.sub2}
\resizebox{5cm}{4.4cm}{\includegraphics{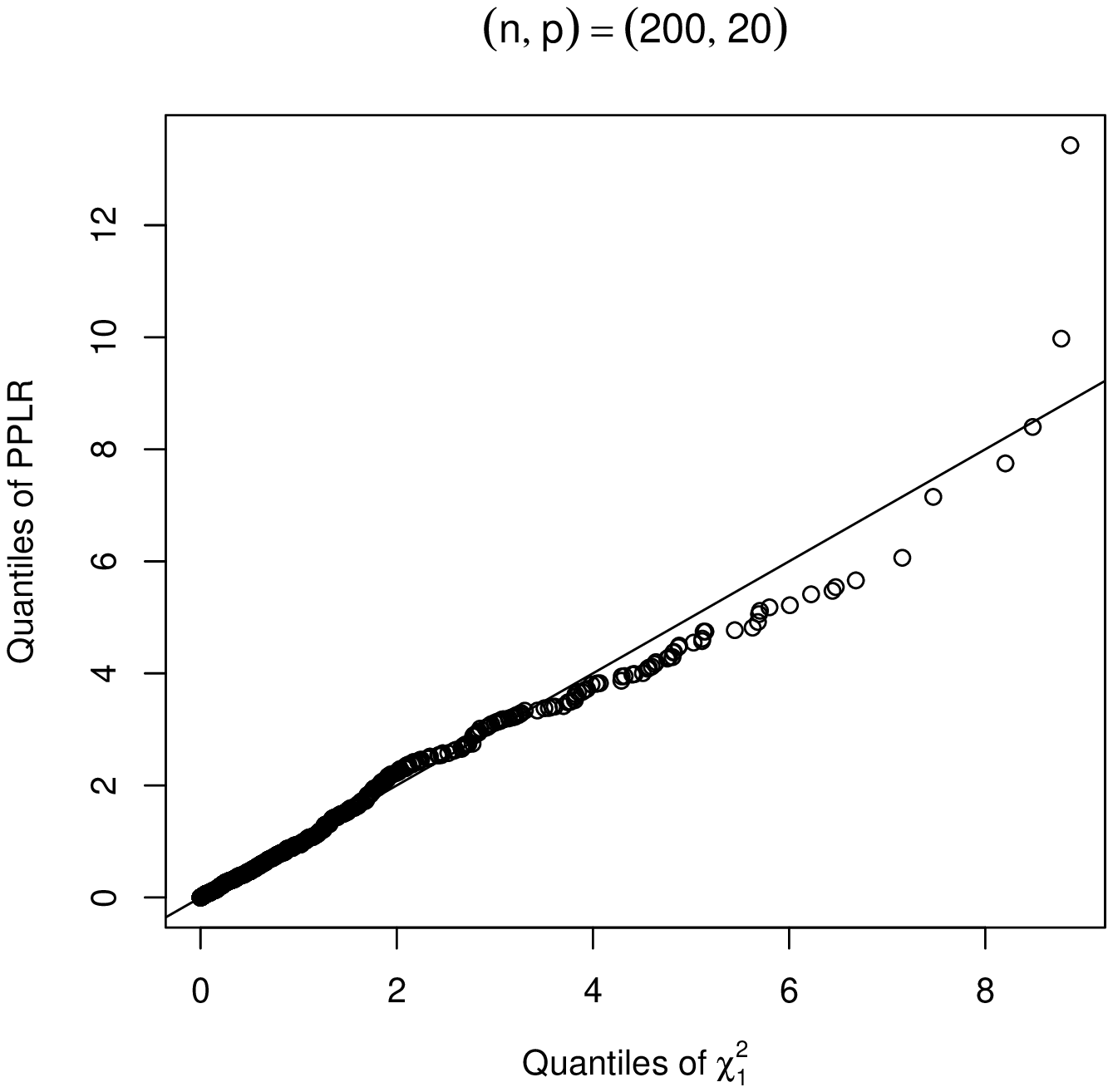}}}
\subfigure[]{
\label{fig1.sub3}
\resizebox{5cm}{4.4cm}{\includegraphics{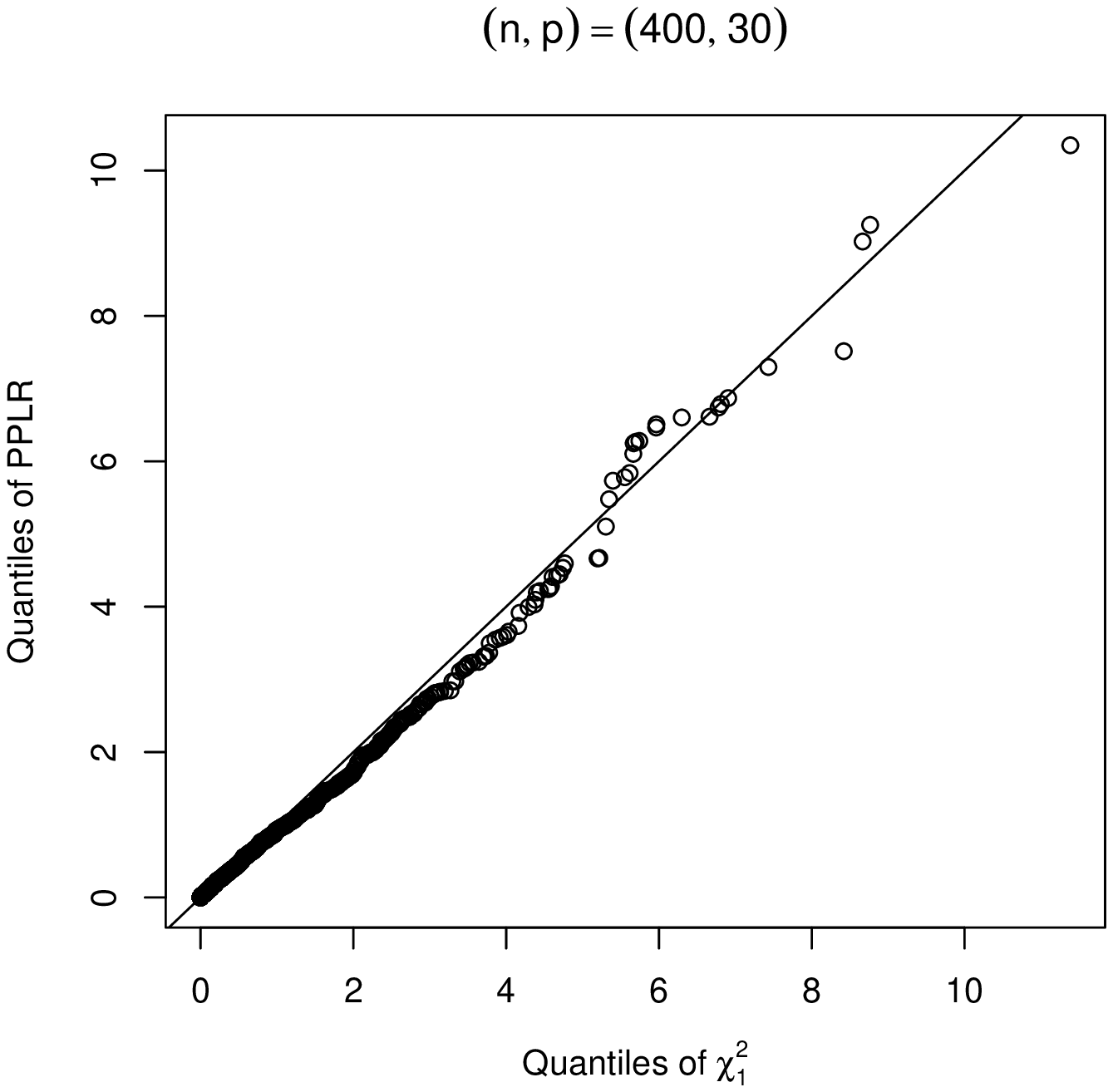}}}
\subfigure[]{
\label{fig1.sub4}
\resizebox{5cm}{4.4cm}{\includegraphics{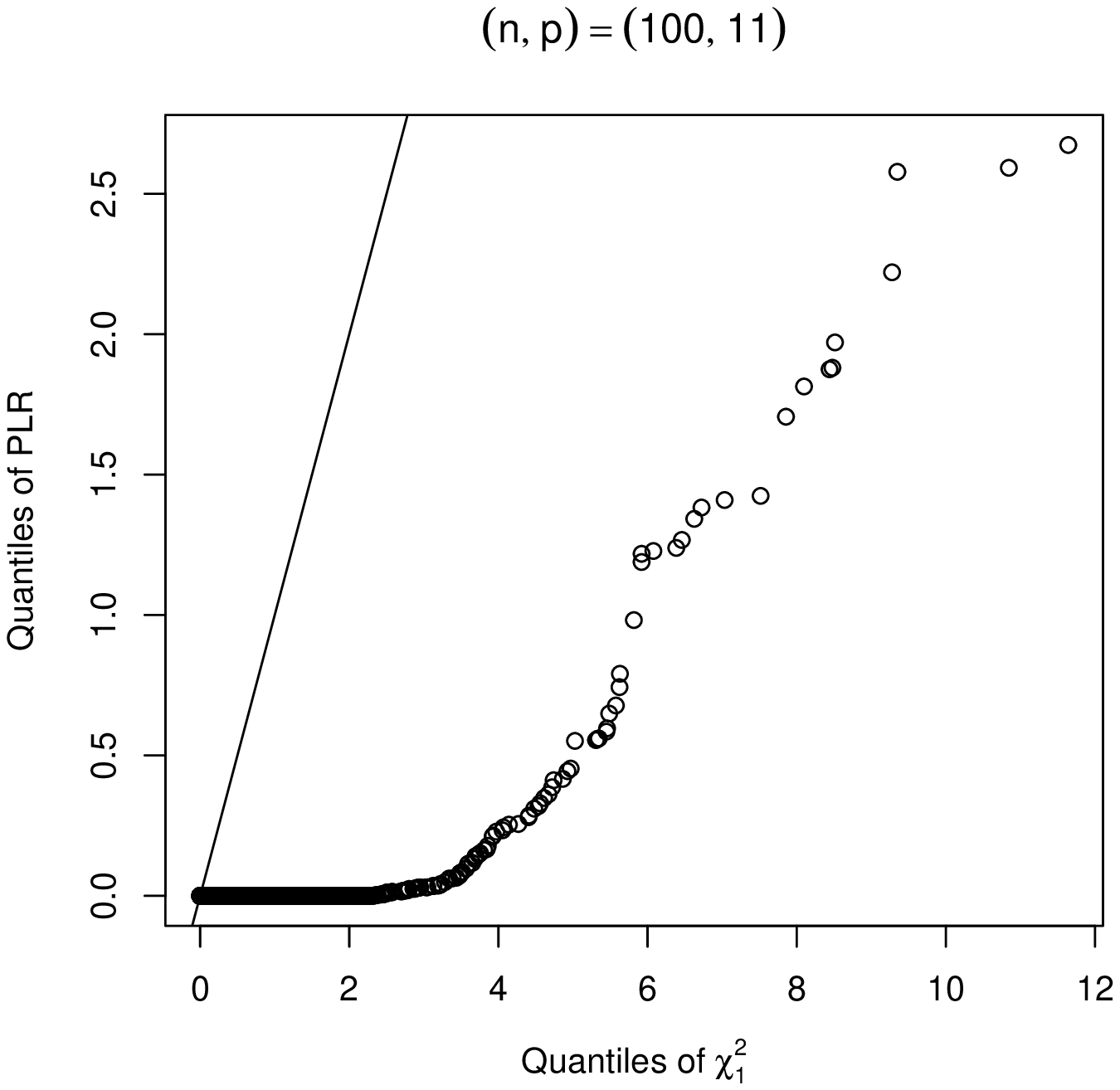}}}
\subfigure[]{
\label{fig1.sub5}
\resizebox{5cm}{4.4cm}{\includegraphics{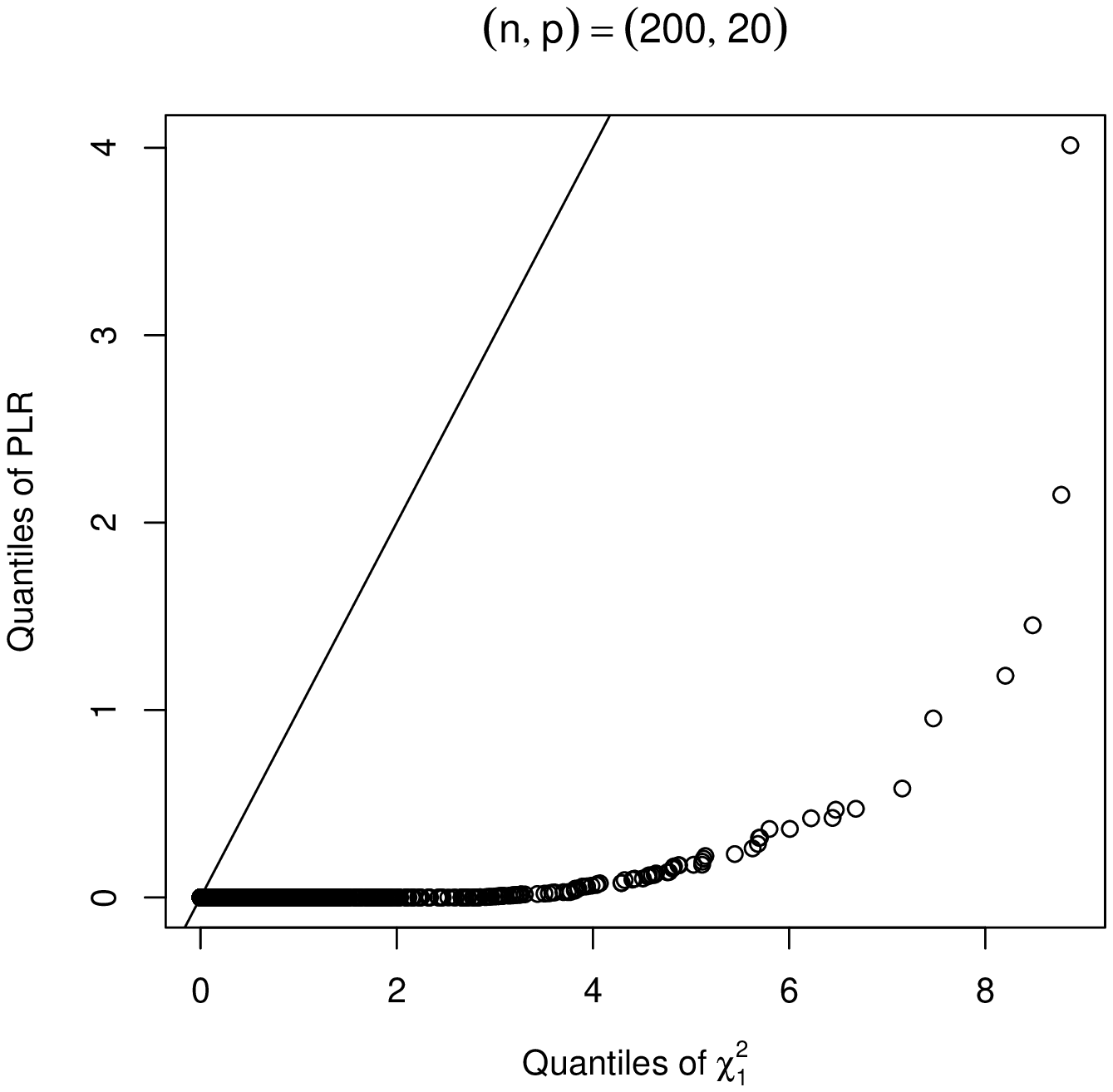}}}
\subfigure[]{
\label{fig1.sub6}
\resizebox{5cm}{4.4cm}{\includegraphics{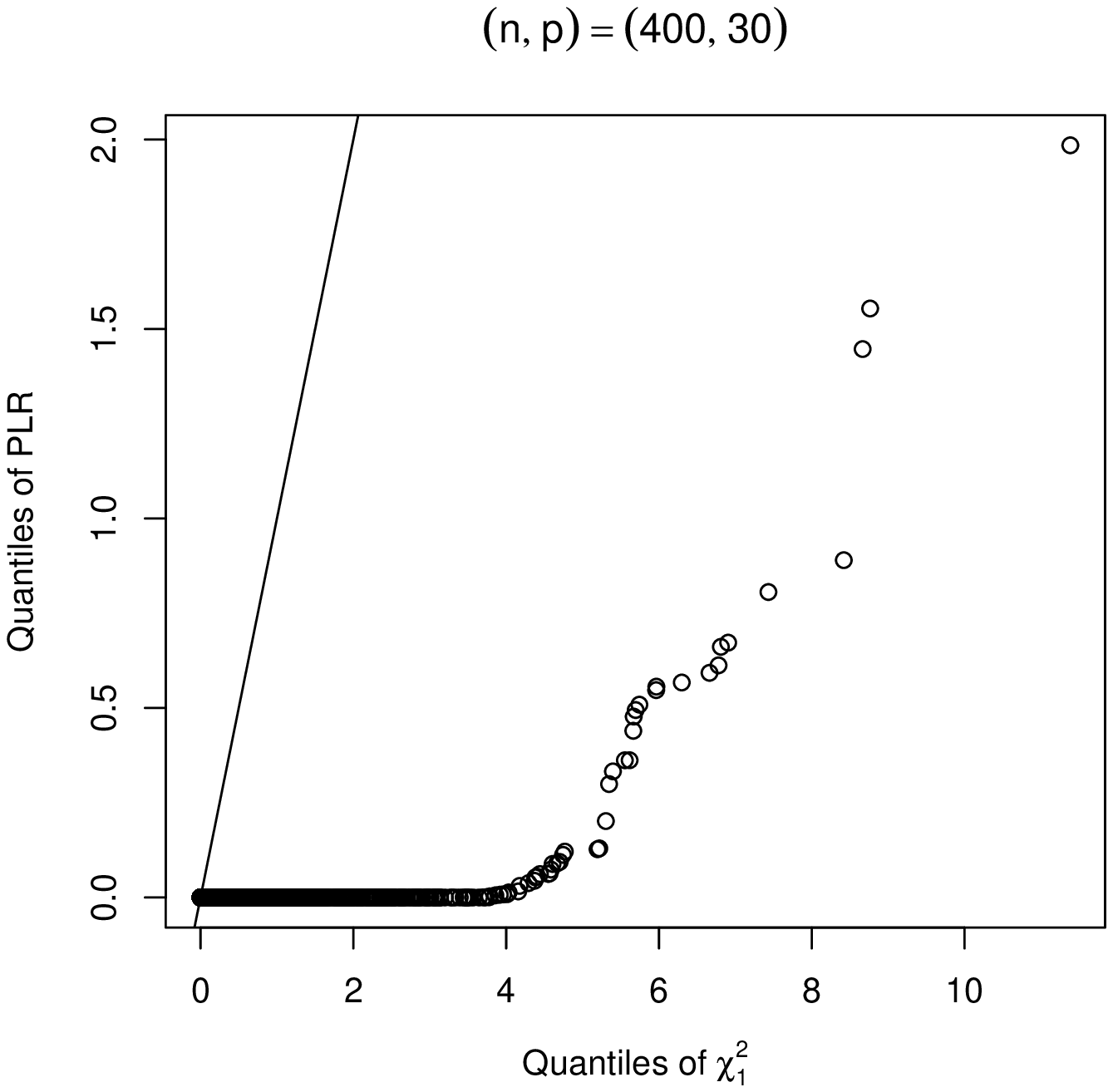}}}
\end{center}
 \caption{QQplots of the partial penalized likelihood ratio (PPLR) test statistics (Top row)
and the penalized likelihood ratio (PLR) test statistics (Bottom row) against
$\chi^2_1$ distribution under null hypothesis $H_0: \beta_{01}=0$ when $(n,p)=(100,11)$ (Left column),
$(n,p)=(200,20)$ (Middle column) and $(n,p)=(400,30)$ (Right column), respectively.}
\label{fig1}
\end{figure}

\begin{figure}{}
\begin{center}
\subfigure[]{
\label{fig2.sub1}
\resizebox{5cm}{4.4cm}{\includegraphics{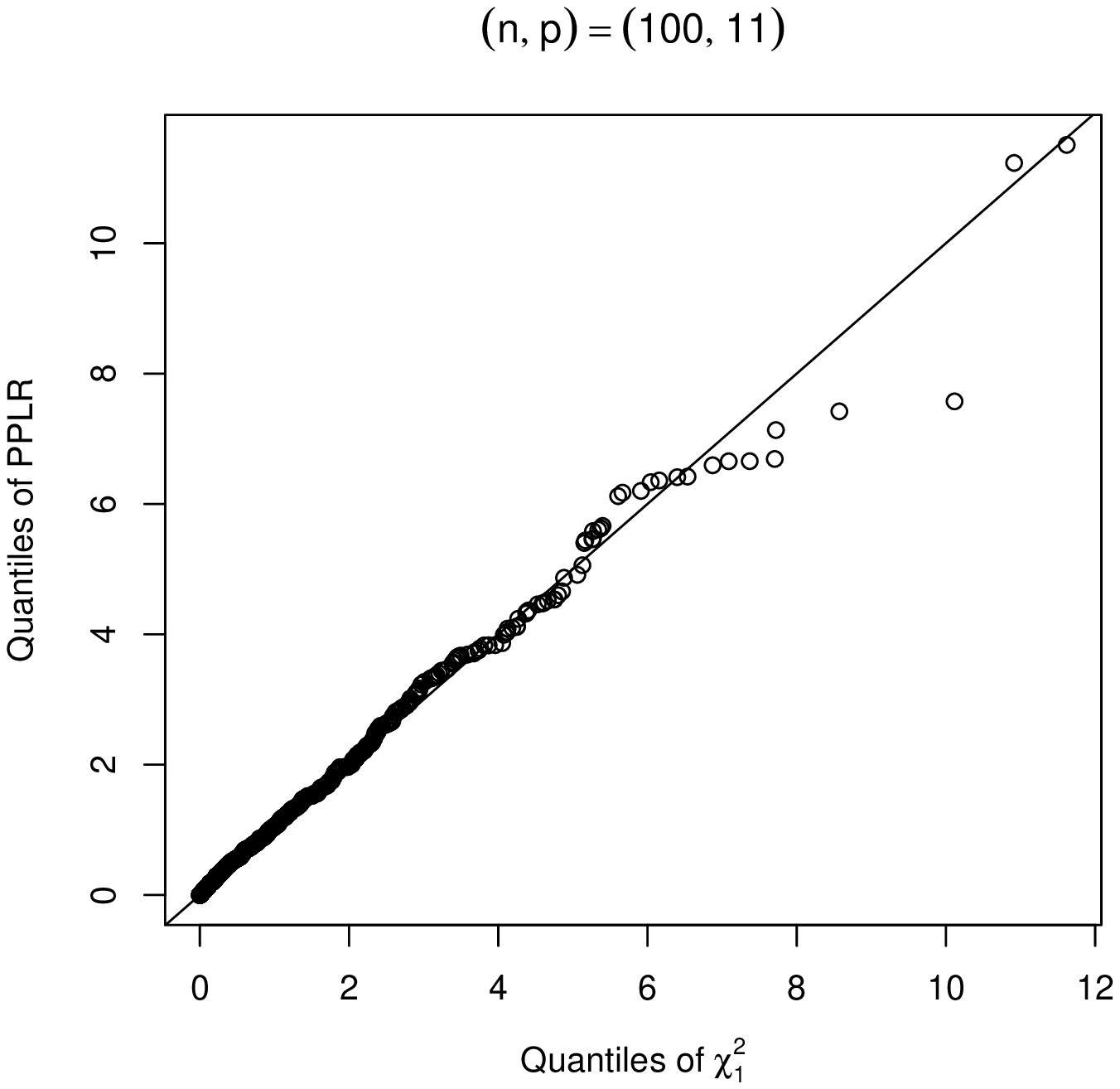}}}
\subfigure[]{
\label{fig2.sub2}
\resizebox{5cm}{4.4cm}{\includegraphics{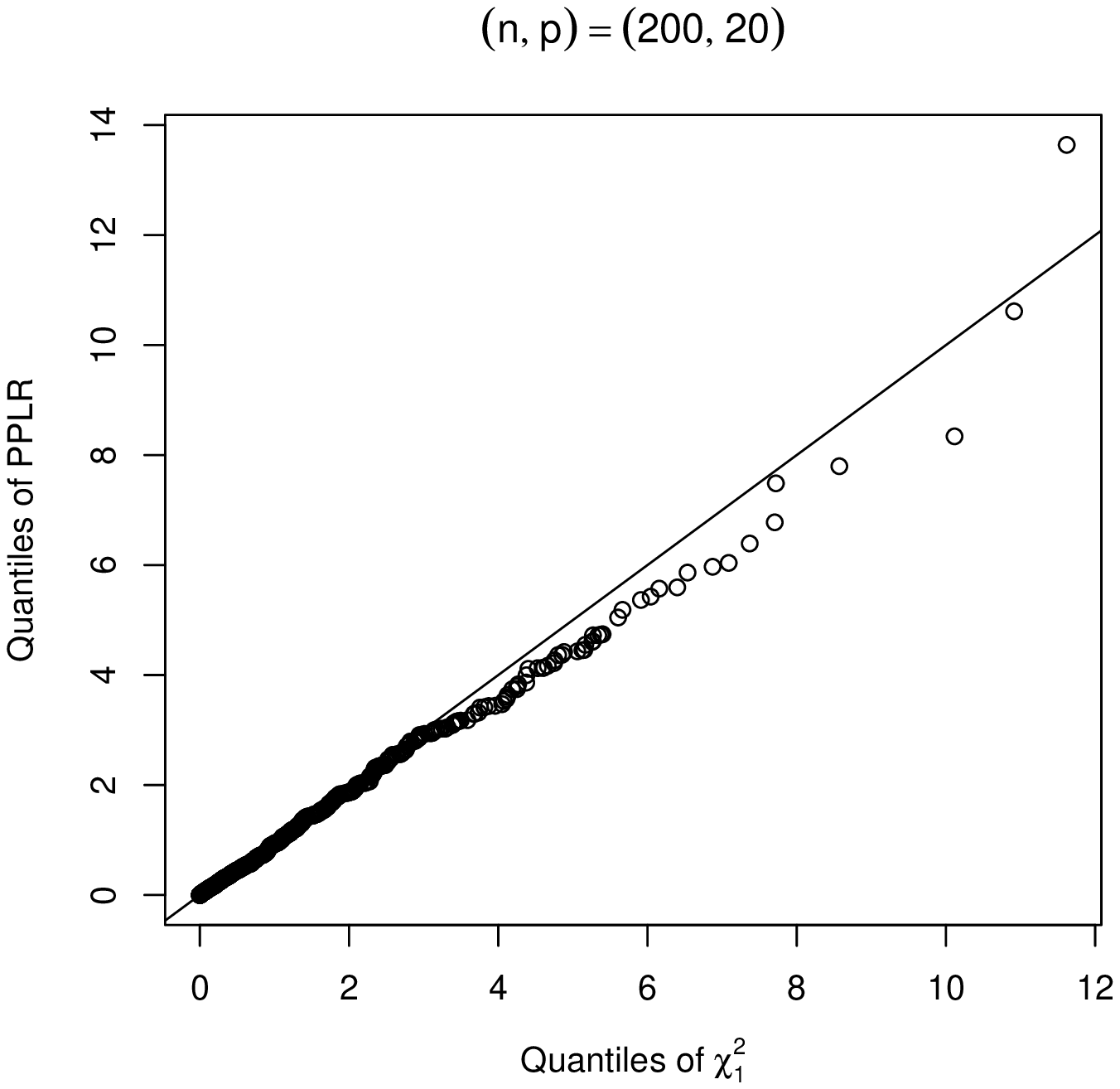}}}
\subfigure[]{
\label{fig2.sub3}
\resizebox{5cm}{4.4cm}{\includegraphics{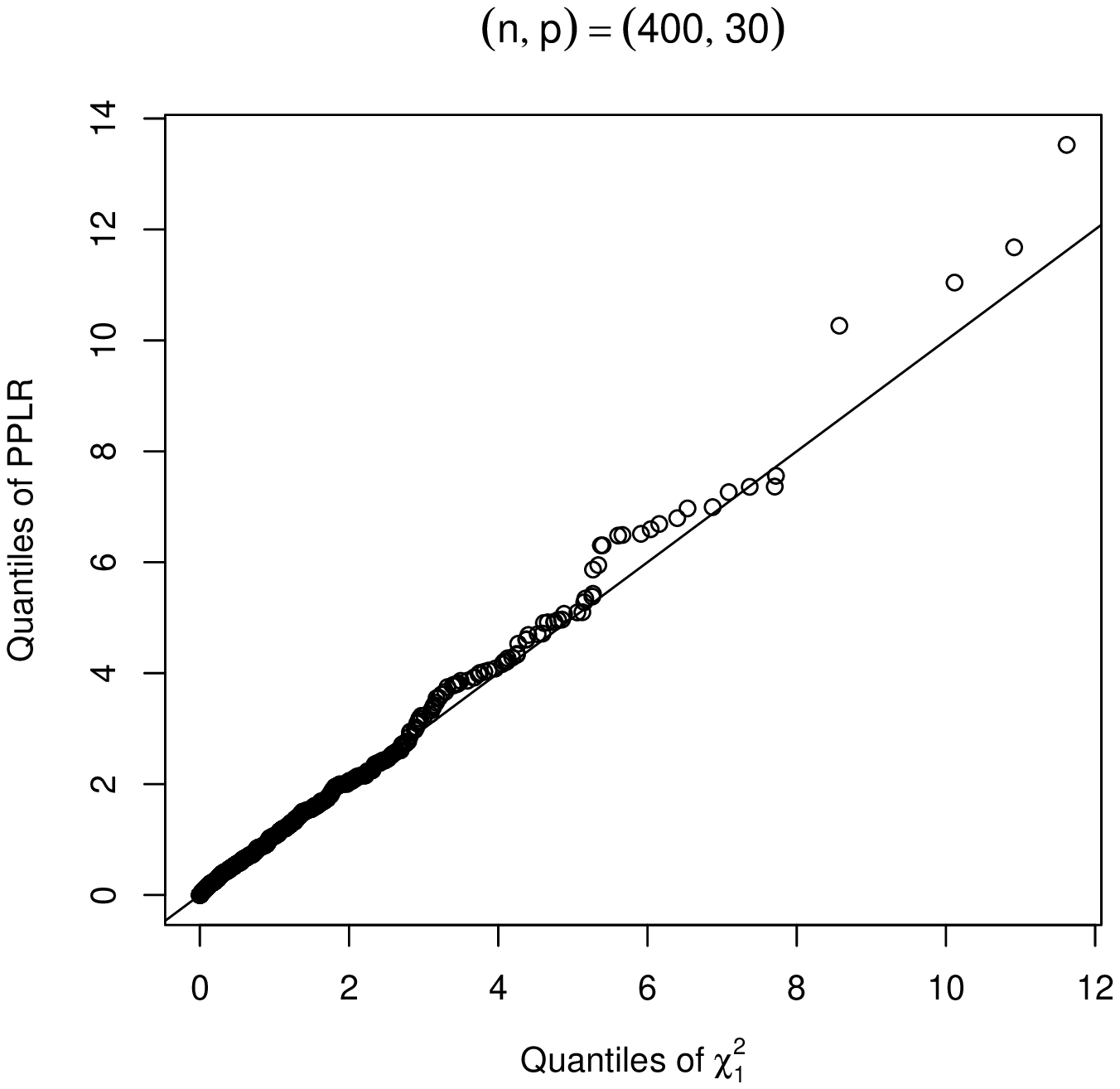}}}
\subfigure[]{
\label{fig2.sub4}
\resizebox{5cm}{4.4cm}{\includegraphics{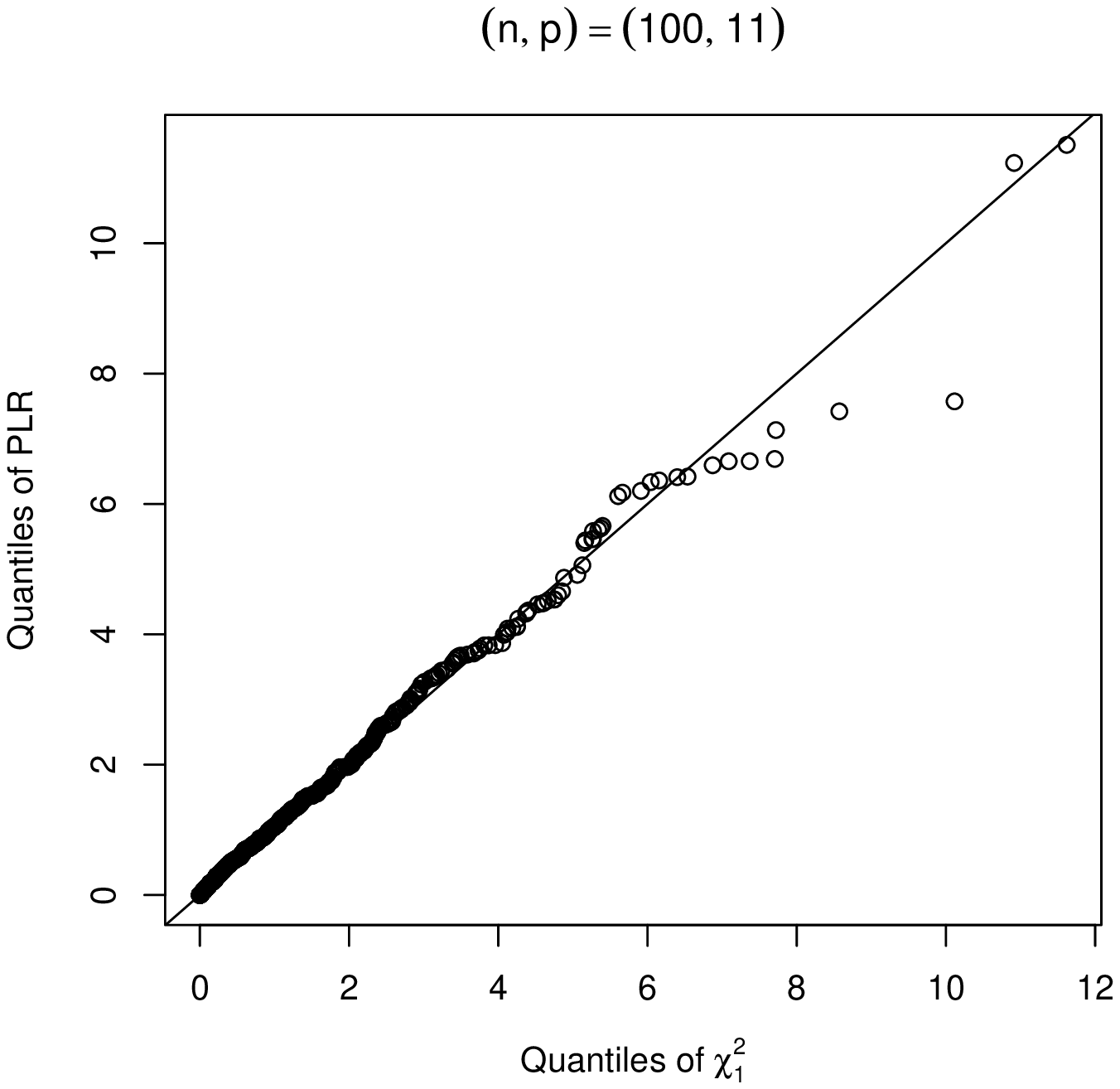}}}
\subfigure[]{
\label{fig2.sub5}
\resizebox{5cm}{4.4cm}{\includegraphics{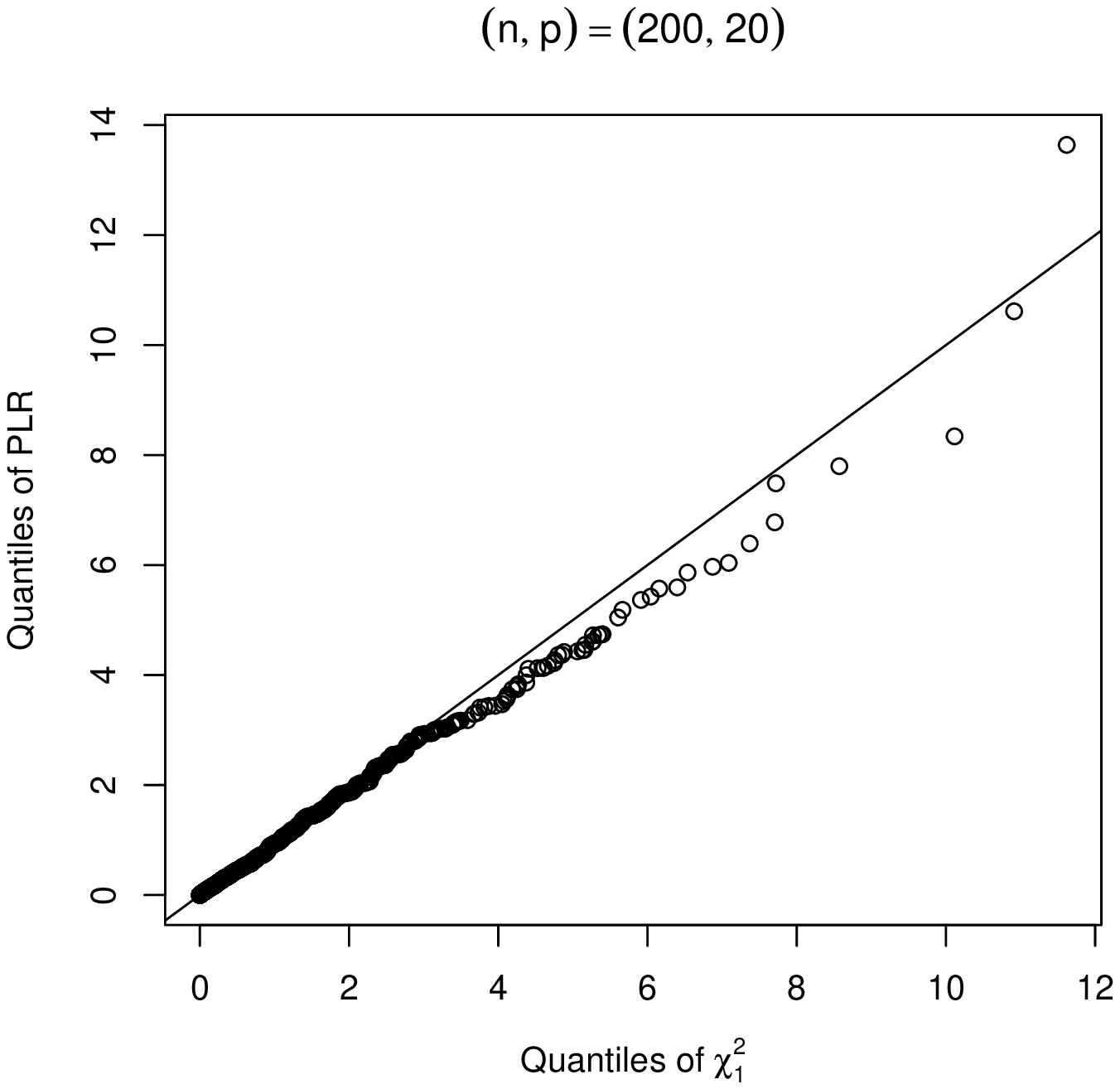}}}
\subfigure[]{
\label{fig2.sub6}
\resizebox{5cm}{4.4cm}{\includegraphics{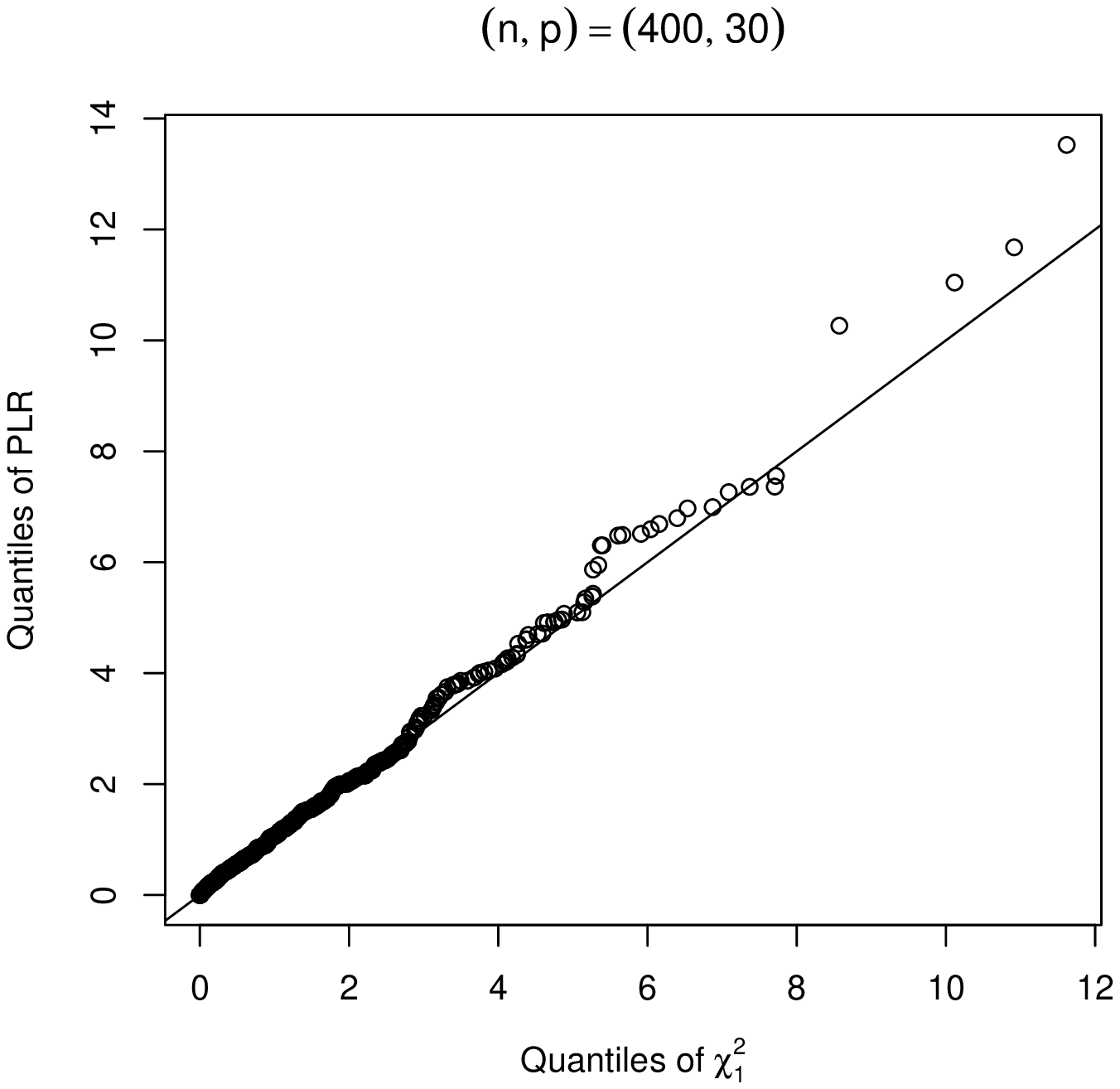}}}
\end{center}
 \caption{QQplots of the partial penalized likelihood ratio (PPLR) test statistics (Top row)
and the penalized likelihood ratio (PLR) test statistics (Bottom row) against
$\chi^2_1$ distribution under null hypothesis $H_0: \beta_{01}=2$ when $(n,p)=(100,11)$ (Left column),
$(n,p)=(200,20)$ (Middle column) and $(n,p)=(400,30)$ (Right column), respectively.}
\label{fig2}
\end{figure}

Meanwhile, from Table \ref{tab2}, we conjecture that under the null
hypothesis $H_0: \beta_{01}=0$, the PLR test
may be not asymptotically chi-squared distributed with
one degree of freedom ($\chi^2_1$), while the conclusion for the
PPLR test still hold. We
demonstrates these results in Figure \ref{fig1} for $n=100, 200$ and
$400$, respectively. Figures \ref{fig1} and \ref{fig2} shows the QQplots of the
PPLR and PLR tests against the nominal $\chi^2_1$
distribution under the null hypothesis $H_0: \beta_{01}=0$ and $H_0: \beta_{01}=2$, respectively. From Figures \ref{fig1} and \ref{fig2}, it follows that the conclusion for the PLR test obtained in \cite{FanPeng2004} only hold
when the null parameter deviates away from zero (For example, $H_0:
\beta_{01}=2$, see Figure \ref{fig2}), and may not hold under the null hypothesis $H_0:
\beta_{01}=0$ (See Figure \ref{fig1}).

For Example \ref{example2}, results in a similar manner to those for Example \ref{example1} are observed, see Tables \ref{tab3} and \ref{tab4}, and from which we can obtain the same conclusions as those for Example \ref{example1}.

\begin{table}[!h]
\begin{center}
\caption{Results for three methods PPL, PL and OL in the
Example \ref{example2} under the true values
$\beta_{0}=(\beta_{01},3,1.5,2,1,0,\cdots,0)^T$, where the first
component $\beta_{01}=\delta n^{-1/2}$ with different $\delta$.
Values shown are means (standard deviations) of each performance
measure over 1000 replicates.}\label{tab3}\vspace{-1cm}
\end{center}
\setlength\tabcolsep{1pt}
\renewcommand{\arraystretch}{1.2}
\begin{center}\scriptsize
\begin{tabular}{ccccccccccccc}
\hline
\multicolumn{2}{c}{Method}&\multicolumn{4}{c}{PPL}&\multicolumn{4}{c}{PL}&\multicolumn{2}{c}{OL}\tabularnewline
\hline
$(n,p)$&$\delta$&\multicolumn{1}{c}{$L_2$-loss}&\multicolumn{1}{c}{$L_1$-loss}&\multicolumn{1}{c}{C}&\multicolumn{1}{c}{IC}&\multicolumn{1}{c}{$L_2$-loss}&\multicolumn{1}{c}{$L_1$-loss}&\multicolumn{1}{c}{C}&\multicolumn{1}{c}{IC}&\multicolumn{1}{c}{$L_2$-loss}&\multicolumn{1}{c}{$L_1$-loss}\tabularnewline
\hline
(100,11)&0.0&$0.220(0.068)$&$0.429(0.144)$&$5.316(0.787)$&$0(0)$&$0.196(0.067)$&$0.352(0.129)$&$6.213(0.850)$&$0.000(0.000)$&$0.338(0.076)$&$0.917(0.223)$\tabularnewline
&1.0&$0.225(0.075)$&$0.443(0.157)$&$5.307(0.773)$&$0(0)$&$0.224(0.067)$&$0.453(0.142)$&$5.266(0.790)$&$0.726(0.446)$&$0.344(0.080)$&$0.932(0.233)$\tabularnewline
&2.0&$0.226(0.069)$&$0.444(0.143)$&$5.264(0.804)$&$0(0)$&$0.256(0.067)$&$0.508(0.150)$&$5.243(0.816)$&$0.330(0.470)$&$0.346(0.077)$&$0.940(0.221)$\tabularnewline
&3.0&$0.224(0.071)$&$0.440(0.150)$&$5.308(0.789)$&$0(0)$&$0.272(0.078)$&$0.526(0.163)$&$5.272(0.795)$&$0.090(0.286)$&$0.342(0.078)$&$0.928(0.226)$\tabularnewline
&4.0&$0.218(0.070)$&$0.430(0.146)$&$5.300(0.816)$&$0(0)$&$0.260(0.088)$&$0.502(0.174)$&$5.298(0.807)$&$0.005(0.071)$&$0.338(0.076)$&$0.916(0.216)$\tabularnewline
(200,20)&0.0&$0.161(0.051)$&$0.329(0.112)$&$14.684(1.108)$&$0(0)$&$0.144(0.049)$&$0.274(0.101)$&$15.570(1.153)$&$0.000(0.000)$&$0.334(0.054)$&$1.236(0.213)$\tabularnewline
&1.0&$0.155(0.048)$&$0.319(0.106)$&$14.582(1.146)$&$0(0)$&$0.154(0.042)$&$0.326(0.097)$&$14.574(1.145)$&$0.743(0.437)$&$0.334(0.055)$&$1.234(0.215)$\tabularnewline
&2.0&$0.159(0.052)$&$0.325(0.111)$&$14.639(1.133)$&$0(0)$&$0.185(0.046)$&$0.376(0.109)$&$14.631(1.131)$&$0.393(0.489)$&$0.335(0.056)$&$1.238(0.218)$\tabularnewline
&3.0&$0.161(0.053)$&$0.329(0.113)$&$14.566(1.146)$&$0(0)$&$0.197(0.056)$&$0.395(0.123)$&$14.510(1.141)$&$0.093(0.291)$&$0.337(0.057)$&$1.247(0.221)$\tabularnewline
&4.0&$0.162(0.048)$&$0.333(0.106)$&$14.572(1.117)$&$0(0)$&$0.196(0.060)$&$0.392(0.124)$&$14.525(1.129)$&$0.019(0.137)$&$0.338(0.054)$&$1.252(0.210)$\tabularnewline
(400,30)&0.0&$0.114(0.034)$&$0.240(0.078)$&$24.134(1.325)$&$0(0)$&$0.103(0.034)$&$0.201(0.072)$&$25.061(1.366)$&$0.000(0.000)$&$0.288(0.039)$&$1.285(0.183)$\tabularnewline
&1.0&$0.114(0.034)$&$0.240(0.074)$&$24.171(1.323)$&$0(0)$&$0.114(0.031)$&$0.245(0.070)$&$24.129(1.311)$&$0.770(0.421)$&$0.289(0.036)$&$1.297(0.173)$\tabularnewline
&2.0&$0.114(0.033)$&$0.239(0.075)$&$24.170(1.303)$&$0(0)$&$0.132(0.029)$&$0.276(0.073)$&$24.149(1.282)$&$0.470(0.499)$&$0.287(0.038)$&$1.290(0.180)$\tabularnewline
&3.0&$0.113(0.033)$&$0.238(0.075)$&$24.168(1.302)$&$0(0)$&$0.142(0.036)$&$0.289(0.081)$&$24.150(1.322)$&$0.121(0.326)$&$0.287(0.038)$&$1.283(0.179)$\tabularnewline
&4.0&$0.113(0.035)$&$0.238(0.080)$&$24.217(1.269)$&$0(0)$&$0.140(0.043)$&$0.285(0.091)$&$24.208(1.245)$&$0.013(0.113)$&$0.286(0.038)$&$1.283(0.181)$\tabularnewline
(800,41)&0.0&$0.080(0.024)$&$0.170(0.056)$&$34.901(1.422)$&$0(0)$&$0.072(0.024)$&$0.143(0.052)$&$35.811(1.471)$&$0.000(0.000)$&$0.234(0.026)$&$1.219(0.146)$\tabularnewline
&1.0&$0.081(0.024)$&$0.171(0.055)$&$34.948(1.383)$&$0(0)$&$0.081(0.022)$&$0.175(0.052)$&$34.943(1.401)$&$0.801(0.399)$&$0.233(0.026)$&$1.208(0.143)$\tabularnewline
&2.0&$0.080(0.024)$&$0.169(0.055)$&$34.942(1.422)$&$0(0)$&$0.093(0.021)$&$0.197(0.053)$&$34.931(1.406)$&$0.442(0.497)$&$0.235(0.026)$&$1.221(0.142)$\tabularnewline
&3.0&$0.080(0.025)$&$0.170(0.055)$&$34.891(1.421)$&$0(0)$&$0.102(0.026)$&$0.209(0.058)$&$34.879(1.404)$&$0.132(0.339)$&$0.235(0.026)$&$1.220(0.144)$\tabularnewline
&4.0&$0.080(0.024)$&$0.171(0.056)$&$34.923(1.395)$&$0(0)$&$0.101(0.030)$&$0.206(0.064)$&$34.909(1.419)$&$0.016(0.126)$&$0.235(0.027)$&$1.226(0.153)$\tabularnewline
\hline
\end{tabular}
\end{center}
\end{table}

\begin{table}[!h]
\begin{center}
\scriptsize \caption{Empirical percentage of rejecting $H_0:
\beta_{01}=0$ for the true values under $H_1:
\beta_{01}=\delta n^{-1/2}$ with different $\delta$ and the sample size $n$ in Example \ref{example2}. The nominal
level is 5\%.}\label{tab4}\vspace{-5mm}
\end{center}
\setlength\tabcolsep{4 pt}
\renewcommand{\arraystretch}{1}
\begin{center}\scriptsize
\begin{tabular}{ccccccccccc}
\hline
$(n,p)$&\multicolumn{1}{c}{Test}&\multicolumn{1}{c}{$\delta=0.0$}&\multicolumn{1}{c}{$\delta=0.5$}&\multicolumn{1}{c}{$\delta=1.0$}&\multicolumn{1}{c}{$\delta=1.5$}&\multicolumn{1}{c}{$\delta=2.0$}
&\multicolumn{1}{c}{$\delta=2.5$}&\multicolumn{1}{c}{$\delta=3.0$}&\multicolumn{1}{c}{$\delta=3.5$}&\multicolumn{1}{c}{$\delta=4.0$}\\
\hline
(100,11)&PPLR&$0.052$&$0.067$&$0.156$&$0.314$&$0.505$&$0.644$&$0.838$&$0.911$&$0.982$\tabularnewline
&PLR&$0.000$&$0.000$&$0.005$&$0.019$&$0.059$&$0.140$&$0.285$&$0.472$&$0.647$\tabularnewline
&LR&$0.050$&$0.063$&$0.149$&$0.301$&$0.485$&$0.620$&$0.816$&$0.903$&$0.970$\tabularnewline
(200,20)&PPLR&$0.064$&$0.069$&$0.179$&$0.288$&$0.502$&$0.680$&$0.857$&$0.941$&$0.970$\tabularnewline
&PLR&$0.000$&$0.001$&$0.004$&$0.011$&$0.058$&$0.134$&$0.252$&$0.439$&$0.621$\tabularnewline
&LR&$0.061$&$0.071$&$0.173$&$0.269$&$0.471$&$0.643$&$0.815$&$0.915$&$0.965$\tabularnewline
(400,30)&PPLR&$0.052$&$0.074$&$0.190$&$0.317$&$0.479$&$0.707$&$0.845$&$0.932$&$0.982$\tabularnewline
&PLR&$0.000$&$0.000$&$0.001$&$0.016$&$0.035$&$0.104$&$0.212$&$0.406$&$0.574$\tabularnewline
&LR&$0.052$&$0.070$&$0.175$&$0.306$&$0.465$&$0.685$&$0.825$&$0.917$&$0.979$\tabularnewline
(800,41)&PPLR&$0.058$&$0.073$&$0.183$&$0.296$&$0.543$&$0.696$&$0.860$&$0.939$&$0.981$\tabularnewline
&PLR&$0.000$&$0.000$&$0.001$&$0.015$&$0.034$&$0.092$&$0.178$&$0.346$&$0.566$\tabularnewline
&LR&$0.054$&$0.072$&$0.178$&$0.284$&$0.517$&$0.669$&$0.848$&$0.930$&$0.979$\tabularnewline
\hline
\end{tabular}
\end{center}
\end{table}

\section{A real data example: Prostate Cancer}\label{sec5}

In this section we illustrate the techniques of our method via an
analysis of Prostate Cancer. The data come from a study by \cite{stamey1989prostateII}, and was analyzed by \cite{HastieTibshirani-114}. They examined
the correlation between the level of prostate-specific antigen and a
number of clinical measures in men who were about to receive a
radical prostatectomy. The sample size is 97 and the variables are
log cancer volume (\textbf{lcavol}), log prostate weight
(\textbf{lweight}), \textbf{age}, log of the amount of benign
prostatic hyperplasia (\textbf{lbph}), seminal vesicle invasion
(\textbf{svi}), log of capsular penetration (\textbf{lcp}), Gleason
score (\textbf{gleason}), and percent of Gleason scores 4 or 5
(\textbf{pgg45}). Among these variables, \textbf{svi} is a binary
variable, and \textbf{gleason} is an ordered categorical variable.
According to the Examples 3.2.1 and 3.3.4 of Hastie et al. (2009),
we know that \textbf{lcavol}, \textbf{lweight} and \textbf{svi} show
a strong relationship with the response \textbf{lpsa}, while
\textbf{age}, \textbf{lcp}, \textbf{gleason} and \textbf{pgg45} are
not significant.  Here, we also fit a linear model to the log of
prostate-specific antigen, \textbf{lpsa}, after first standardizing
the predictors to have unit variance and centering the response to
have zero mean.
\begin{equation}\label{model}
  \textbf{lpsa}=\beta_1 \textbf{lcavol}+\beta_2
    \textbf{lweight}+\beta_3
    \textbf{age}+\beta_4 \textbf{lbph}+\beta_5 \textbf{svi}+\beta_6 \textbf{lcp}+\beta_7
    \textbf{gleason}+\beta_8 \textbf{pgg45}+\varepsilon.
\end{equation}
We are interested in the significance of each predictor, which
leads to the null hypothesis $H_{0,j}:\ \beta_j=0$ for the
individual $j$th predictor, where $j=1, 2, \cdots, 8$.

We applied the ordinary least-squares fit (LS), the penalized
likelihood method (SCAD-PL) and the partial penalized likelihood
method (SCAD-PPL), and the estimated coefficients and their multiple
$R^2$ are summarized in the Table \ref{tab5} with $\beta_5$ non-penalized in the SCAD-PPL method, from which we can see that over
$60\%$ of the \textbf{lpsa} variation can be explained by the
variables that we use, and the results are consistent with the
analysis of Examples 3.2.1 and 3.3.4 of \cite{HastieTibshirani-114}.

\begin{table}[!h]
\begin{center}
\scriptsize \caption{ Estimates for Prostate Cancer
data}\label{tab5}\vspace{-0.4cm}
\end{center}
\setlength\tabcolsep{10 pt}
\renewcommand{\arraystretch}{1}\scriptsize
\begin{center}
\begin{tabular}{cccc}
\hline
\multicolumn{1}{c}{Variable}&\multicolumn{1}{c}{LS}&\multicolumn{1}{c}{SCAD-PL}&\multicolumn{1}{c}{SCAD-PPL}\\
\hline
lcavol&$ 0.6651$&$0.6662$&$0.6324$\\
lweight&$ 0.2665$&$0.2415$&$0.2312$\\
age&$-0.1582$&$0.0000$&$0.0000$\\
lbph&$ 0.1403$&$0.0089$&$0.0202$\\
svi&$ 0.3153$&$0.2111$&$0.2786$\\
lcp&$-0.1483$&$0.0000$&$0.0000$\\
gleason&$ 0.0355$&$0.0000$&$0.0000$\\
pgg45&$ 0.1257$&$0.0000$&$0.0000$\\
$R^2$& $0.6633$&$0.6052$&$0.6208$\\
\hline
\end{tabular}
\end{center}
\end{table}

For the null hypothesis $H_{0,j}$, Table \ref{tab6} summarizes p-values of
testing results, using unpenalized, full penalized and partial
penalized versions of the likelihood ratio test. At a nominal
significance level 0.05, the test results of PPLR method shows that
three predictors \textbf{lcavol}, \textbf{lweight} and \textbf{svi}
are significant, which is consistent with variable selection results
in Table \ref{tab5}. However, the test results of PLR for the predictor
\textbf{svi} contracts (\textbf{svi} is insignificant). Again, this
phenomena shows that full penalized likelihood ratio test can not
distinguish the nonzero component that is near zero.

\begin{table}[!h]
\begin{center}
\scriptsize \caption{ P-values for each individual predictor of
Prostate Cancer data}\label{tab6}\vspace{-0.4cm}
\end{center}
\setlength\tabcolsep{10 pt}
\renewcommand{\arraystretch}{1}\scriptsize
\begin{center}
\begin{tabular}{cccc}
\hline
\multicolumn{1}{c}{Method}&\multicolumn{1}{c}{LR}&\multicolumn{1}{c}{SCAD-PLR}&\multicolumn{1}{c}{SCAD-PPLR}\\
\hline lcavol&$0.0000$&$0.0000$&$0.0000$\\
lweight&$0.0303$&$0.1373$&$0.0161$\\
age&$0.1800$&$1.0000$&$0.4641$\\
lbph&$0.2427$&$0.9383$&$0.2257$\\
svi&$0.0272$&$0.2070$&$0.0241$\\
lcp&$0.4092$&$1.0000$&$0.9748$\\
gleason&$0.8248$&$1.0000$&$0.5307$\\
pgg45&$0.4751$&$1.0000$&$0.3577$\\ \hline
\end{tabular}
\end{center}
\end{table}

\section{Conclusion and discussion}

Based on the idea of partial penalization, this paper propose a consistent test, called the partial penalized likelihood ratio test for the hypothesis problem (\ref{hypothesis}) in the framework that $p$ diverging with $n$, establishing that the proposed test converges in distribution to $\chi^2_d$ under $H_0$ (See Theorem \ref{theorem1}) and $\chi^2_d(\gamma)$ with the noncentral parameter $\gamma$ depending on $\delta$ under the local alternatives of order $n^{-1/2}$ (See Theorem \ref{theorem2}), respectively. Meanwhile, the proposed partial penalized likelihood method also can perform variable selection for the sparse parameter $\beta_2$, keeping the oracle property. In this sense, our proposed a consistent test performs as well as the OLR, and the PPL method is also capable of selecting important variables as PL method, achieving better predictability and estimation accuracy. Overall, the main contribution of this paper is to propose the idea of partial penalization as well as a consistent test for (\ref{hypothesis}), demonstrating its promising advantage in variable selection and hypothesis testing. And we also conduct some numerical simulations and an analysis of Prostate Cancer data to confirm our theoretical findings and demonstrate the promising performance of the proposed method.

It is noted in the simulation that our proposed test performs as
well as the classical likelihood ratio test in term of size and
power. Yet, the benefit of our proposed method compared with the
classical likelihood method is to conduct variable selection for the
rest sparse parameter, obtaining better estimation accuracy. This
is, our proposed method can perform hypothesis testing and variable
selection simultaneously.

In the present paper is assumed that the position of $d$ parameters
of interest is known, that is the proposed partial penalized
likelihood method only applies to the proposed null hypothesis of
the form (\ref{hypothesis}). This is partly
motivated by some prior knowledge or else that the parameter of
interest is pre-specified. However, when the position of $d$
parameters of interest is unknown, the proposed method may be not
applicable. For example, the null hypothesis is all components of
$p$-dimensional parameter is zero while the alternative hypothesis
is the number of nonzero components is $d$ at most, where $d$ is
known and $d\ll p$. How to test it via the proposed partial
penalized likelihood method? These question deserves our further
study, and has been in progress, but is beyond the scope of the
current paper.

\section*{Acknowledgements}
Cui's research was supported in part by the National Natural Science Foundation of China (Grant Nos. 11071022, 11028103, 11231010, 11471223), the Key project of Beijing Municipal Education Commission (Grant No. KZ201410028030) and the Foundation of Beijing Center for Mathematics and Information Interdisciplinary Sciences.
And we are grateful to Assistant Prof. PingShou Zhong for constructive comments.

\section*{Appendix-Proofs of theorems}

\setcounter{section}{1}
\setcounter{equation}{0}
\def\theequation{A.\arabic{equation}}
\def\thesection{A\arabic{section}}


To establish Theorems \ref{theorem1} and \ref{theorem2}, we present the following lemmas as well as the regularity conditions similar as \cite{FanPeng2004} here. The conditions that imposed on the
likelihood function are:

\no{\bf(A)} For every $n$ observations $\{V_{i}\}_{i=1}^n$ are independent and identically distributed with the
probability density $f(V_{1}, \beta_0)$, which has a common
support, and the model is identifiable. Furthermore, the first and
second derivatives of the likelihood function satisfy equations
$E_{\beta_0}\{\frac{\partial \log f(V_{1}, \beta_0)}{\partial
\beta_{j}}\}=0  \  \mbox{for}\  j=1, 2, \cdots, p$, and
$E_{\beta_0}\{\frac{\partial \log f(V_{1}, \beta_0)}{\partial
\beta_{j}}\frac{\partial \log f(V_{1}, \beta_0)}{\partial
\beta_{k}}\}=-E_{\beta_0}\{\frac{\partial^2 \log f(V_{1},
\beta_0)}{\partial \beta_{j}\partial \beta_{k}}\}$.

\no{\bf(B)} The Fisher information matrix
$I(\beta_0)=E[\{\frac{\partial \log f(V_{1},
\beta_0)}{\partial \beta}\}\{\frac{\partial \log f(V_{1},
\beta_0)}{\partial \beta}\}^{T}]$ satisfies conditions
$$0<C_1<\lambda_{\mbox{min}}\{I(\beta_0)\}\leq\lambda_{\mbox{max}}\{I(\beta_0)\}<C_2<\infty\
\ \ \mbox{for all} \ n,$$ and for $1\leq j, k\leq p$,
$E_{\beta_0}\{\frac{\partial \log f(V_{1}, \beta_0)}{\partial
\beta_{j}}\frac{\partial \log f(V_{1}, \beta_0)}{\partial
\beta_{k}}\}^2<C_{3}<\infty$ and $E_{\beta_0}\{\frac{\partial^2
\log f(V_{1}, \beta_0)}{\partial \beta_{j}\partial
\beta_{k}}\}^2<C_4<\infty$.

\no{\bf(C)} There is a large enough open subset $\omega$ of
$\Omega\subset {\bf R}^{p}$ which contains the true parameter point
$\beta_0$, such that for almost all $V_{i}$ the density admits all
third derivatives $\partial f(V_{1}, \beta)/\partial
\beta_{j}\beta_{k}\beta_{l}$ for all $\beta \in \omega$.
Furthermore, there are functions $M_{jkl}$ such that
$|\frac{\partial^3 }{\partial
\beta_{j}\partial\beta_{k}\partial\beta_{l}}\log f(V_{1},
\beta)|\leq M_{jkl}(V_{1})$ for all $\beta \in \omega$, and
$E_{\beta}\{M^2_{jkl}(V_{1})\}<C_5<\infty$ for all $p, n$ and
$j, k, l$.

The aforementioned conditions are similar as in \cite{FanPeng2004}, under conditions (A) and (C), the second and fourth moments of the
likelihood function are imposed. The information matrix of the
likelihood functions is assumed to be positive definite, and its
eigenvalues are uniformly bounded, which is a common assumption in
the high-dimensional setting. These conditions are stronger that
those of the usual asymptotic likelihood theory, but they facilitate
the technical derivations.

Let $a_{n}=\max_{(d+1)\leq j\leq
p}\{p'_{\lambda}(|\beta_{j0}|), \beta_{j0}\neq 0\}$ and
$b_{n}=\max_{(d+1)\leq j\leq p}\{p''_{\lambda}(|\beta_{j0}|),
\beta_{j0}\neq 0\}$. Then we need to place the following conditions
on the penalty functions:

\no(D) $\lim\inf_{n\rightarrow\infty}\lim\inf_{\theta\rightarrow
0^{+}}p'_{\lambda}(\theta)/\lambda>0$;

\no(E) $a_{n}=O(n^{-1/2})$; ($E'$) $a_n=o(1/\sqrt{np})$;

\no(F) $b_n\rightarrow 0$ as $n\rightarrow\infty$; ($F'$)
$b_n=o(1/\sqrt{p})$;

\no(G) there are constants $C$ and $D$ such that, when $\theta_1,
\theta_{2}> C\lambda_n,
|p''_{\lambda}(\theta_1)-p''_{\lambda}(\theta_2)| \leq
D|\theta_1-\theta_2|$.


\no(H) The nonzero components $\beta_{(d+1)0}, \cdots,
\beta_{(d+s)0}$ satisfy $\min_{d+1\leq j\leq
d+s}\frac{|\beta_{j0}|}{\lambda}\rightarrow\infty$, as
$n\rightarrow\infty$.

Condition (H) can be viewed as ``beta-min" condition, and it states
that the weakest signal should dominate the penalty parameter
$\lambda$, which is routinely made to ensure the recovery of
signals, and is reasonable because otherwise the noise is too
strong. And this also in line with condition imposed in \cite{FanPeng2004}. Given condition (H), all of conditions (D)-(G) are satisfied
by the SCAD penalty, as $a_n=0$ and $b_n=0$ when $n$ is large
enough.

Denote
$\tilde{\Sigma}_{\lambda}=\left(
                         \begin{array}{cc}
                           0_{d\times d} & 0_{d\times s} \\
                           0_{s\times d} & \Sigma_{\lambda}\\
                         \end{array}
                       \right),\ \
                       \tilde{b}_{n}=(0_{d},b^{T}_n)^{T}
$, where $\Sigma_{\lambda}=\mbox{diag}\{p''_{\lambda}(|\beta_{(d+1)0}|), \cdots,
p''_{\lambda}(|\beta_{(d+s)0}|)\}$ and
$b_n=\{p'_{\lambda}(|\beta_{(d+1)0}|)\mbox{sgn}(\beta_{(d+1)0}),
\cdots,
p'_{\lambda}(|\beta_{(d+s)0}|)\mbox{sgn}(\beta_{(d+s)0})\}^{T}$.

\no\textbf{Lemma 1.}\label{lemma1} \textsl{(Existence of partial
penalized likelihood estimator) Suppose that pdf
$f(V,\beta_0)$ satisfies conditions (A)-(C), and the penalty
function $p_{\lambda}(\cdot)$ satisfies conditions (E)-(G). If
$p^4/n\rightarrow 0$ as $n\rightarrow\infty$, then there exists a
local maximizer $\hat{\beta}$ of $PQ_n(\beta|V)$ such that
$\|\hat{\beta}-\beta_0\|=O_p(\sqrt{p}(n^{-1/2}+a_n))$.}

\no\textbf{Lemma 2.} \label{lemma2} \textsl{(Oracle property) Under
conditions (A)-(H), if $\lambda\rightarrow 0$ and
$\sqrt{n/p}\lambda\rightarrow \infty$ and $p^5/n\rightarrow 0$
as $n\rightarrow\infty$, then, with probability tending to 1, the
root-$(n/p)$-consistent nonconcave partial penalized likelihood
estimator $\hat{\beta}=(\hat{\beta}^{\mathcal{D}T},
\hat{\beta}^{\mathcal{I}T})^{T}$ in Lemma 1 must satisfy:
\begin{itemize}
  \item [](i) Sparsity: $\hat{\beta}^{\mathcal{I}}=0$.
  \item [](ii) Asymptotic normality:
$$\sqrt{n}A_nI^{-1/2}_{1}(\beta^{\mathcal{D}}_{0})(I_{1}(\beta^{\mathcal{D}}_{0})+\tilde{\Sigma}_{\lambda})\{(\hat{\beta}^{\mathcal{D}}-\beta^{\mathcal{D}}_{0})+(I_{1}(\beta^{\mathcal{D}}_{0})+\tilde{\Sigma}_{\lambda})^{-1}\tilde{b}_n\} \rightarrow N(0, G),$$
where $I_{1}(\beta^{\mathcal{D}}_{0})=I_{1}(\beta^{\mathcal{D}}_{0},0)$, the
Fisher information knowing $\beta^{\mathcal{I}}=0$, and $A_n$ is a
$q\times (d+s)$ matrix such that $A_nA^{T}_n\rightarrow G$, and
$G$ is a $q\times q$ nonnegative symmetric matrix.
\end{itemize}}

\no{\bf Proof.}\quad
With the working penalty function
$\tilde{p}_{\lambda}(\cdot)$, the proof of Lemmas 1 and 2 follows
the arguments in \cite{FanPeng2004}'s paper, we omit it here.

Let $\Omega_0=\{(\nu_1, \cdots, \nu_{p-d}):
(\beta^T_{01},\nu^{T})^{T}\in \Omega\}$ with $\nu=(\nu^{\mathcal{D}T},
\nu_n^{\mathcal{I}T})^{T}$ ranges
through an open subset of ${\bf R}^{p-d}$, where $\nu^{\mathcal{D}}$
is a $s\times 1$ vector and $\nu^{\mathcal{I}}$ is a $(p-d-s)\times
1$ vectors. The specification of $\Omega_0$ may equivalently be
given as a transformation $\beta_{j}=g_j(\nu)$, where
$g(\nu)=(g_1(\nu), \cdots, g_p(\nu))^{T}$ with the first $d$ components
$\{g_1(\nu),\ldots, g_d(\nu)\}^T=\beta_{01}$, and the remaining $p-d$ components
$g_j(\nu)=\nu_{j-d}$ for $(d+1)\leq j\leq p$. And denote $\nu_0$ is the true value of $\nu$.
Under $H_0$, it follows that
$\beta_0=g(\nu_{0})$. Thus under $H_0$, the partial penalized
likelihood estimator $\tilde{\beta}=g(\hat{\nu})$ is also the
local maximizer $\hat{\nu}$ of the problem
$PQ_{n}(g(\hat{\nu})|V)=\max_{\nu}PQ_n(g(\nu)|V)$.
Note that the first order partial derivatives of function $g$ as
$C=[\frac{\partial g_i}{\partial \nu_{j}}]_{p\times
(p-d)}=\left(
                                           \begin{array}{c}
                                             0_{d\times (p-d)} \\
                                             I_{p-d} \\
                                           \end{array}
                                         \right)
$, and let $D=\left(
                \begin{array}{c}
                 0_{d\times s} \\
                 I_{s} \\
                \end{array}
              \right)
$ be a sub-matrix of $C$.

\no\textbf{Lemma 3.} \label{lemma3} \textsl{Under the condition of Theorem \ref{theorem1}
and the null hypothesis $H_0$, we have
\begin{eqnarray*}
  \hat{\beta}^{\mathcal{D}}-\beta^{\mathcal{D}}_{0} &=& n^{-1}I^{-1}_{1}(\beta^{\mathcal{D}}_{0})\nabla L_n(\beta^{\mathcal{D}}_{0})+o_p(n^{-1/2}), \\
 \tilde{\beta}^{\mathcal{D}}-\beta^{\mathcal{D}}_{0} &=& n^{-1}D(D^{T}I_{1}(\beta^{\mathcal{D}}_{0})D)^{-1}D^{T}\nabla
  L_n(\beta^{\mathcal{D}}_{0})+o_p(n^{-1/2}).
\end{eqnarray*}}
\no{\bf Proof.}\quad
We need only prove the second equation. The first
equation can be show in the same manner. Following the steps of the
proof of Lemma 2, it follows that under $H_0$,
\begin{equation}\label{aeq1}
  (I^{\nu}_1(\nu^{\mathcal{D}}_{0})+\Sigma_{\lambda})(\hat{\nu}^{\mathcal{D}}-\nu^{\mathcal{D}}_{0})-b_n=n^{-1}\nabla
L_n(g(\nu^{\mathcal{D}}_{0},0))+o_p(n^{-1/2}),
\end{equation}
where
$I^{\nu}_1(\nu^{\mathcal{D}}_{0})=I^{\nu}(\nu^{\mathcal{D}}_{0},0)$ with
$I^{\nu}(\nu_{0})$ is the information matrix for the
$\nu$-formulation of model. For $\beta_0=g(\nu_0)$,
$I^{\nu}_1(\nu^{\mathcal{D}}_{0})=D^{T}I_1(\beta^{\mathcal{D}}_{0})D$,
$\nabla L_n(g(\nu^{\mathcal{D}}_{0},0))=D^{T}\nabla
L_n(\beta^{\mathcal{D}}_{0})$, and $\tilde{\Sigma}_\lambda=D\Sigma_\lambda D^T$, we have
\begin{equation}\label{aeq2}D^{T}(I_{1}(\beta^{\mathcal{D}}_{0})+\tilde{\Sigma}_{\lambda})D(\hat{\nu}_n^{\mathcal{D}}-\nu^{\mathcal{D}}_{0})-b_n=n^{-1}D^{T}\nabla
L_n(\beta^{\mathcal{D}}_{0})+o_p(n^{-1/2}).\end{equation} By the conditions
$a_n=o(1/\sqrt{np})$, and $s\leq p$, we
have $\|b_{n}\|\leq \sqrt{s}
a_n=o_p(1/\sqrt{n})$. On the other hand, by condition
$b_n=o_p(1/\sqrt{p})$,
$\|D^{T}\tilde{\Sigma}_{\lambda}D(\hat{\nu}^{\mathcal{D}}-\nu^{\mathcal{D}}_{0})\|=\|\Sigma_{\lambda}(\hat\nu^{\mathcal{D}}-\nu^{\mathcal{D}}_{0})\|\leq
b_n\|\hat\nu^{\mathcal{D}}-\nu^{\mathcal{D}}_{0}\|=o_p(1/\sqrt{n})$. It follows that
$D^{T}I_{1}(\beta^{\mathcal{D}}_{0})D(\hat{\nu}^{\mathcal{D}}-\nu^{\mathcal{D}}_{0})=n^{-1}D^{T}\nabla
L_n(\beta^{\mathcal{D}}_{0})+o_p(n^{-1/2})$. As
$D^{T}I_{1}(\beta^{\mathcal{D}}_{0})D$ is $s\times s$ sub-matrix of
the fisher matrix $I(\beta_{0})$, and by condition (B), if follows that
$$\tilde{\beta}^{\mathcal{D}}-\beta^{\mathcal{D}}_{0}=D(\hat{\nu}^{\mathcal{D}}-\nu^{\mathcal{D}}_{0})
= n^{-1}D(D^{T}I_{1}(\beta^{\mathcal{D}}_{0})D)^{-1}D^{T}\nabla
  L_n(\beta^{\mathcal{D}}_{0})+o_p(n^{-1/2}).$$

\no\textbf{Lemma 4.}\label{lemma4} \textsl{Under the condition of Theorem \ref{theorem1}
and the null hypothesis $H_0$, we have
$$PQ_n(\hat{\beta}^{\mathcal{D}}|V)-PQ_n(\tilde{\beta}^{\mathcal{D}}|V)=\frac{n}{2}(\hat{\beta}^{\mathcal{D}}-\tilde{\beta}^{\mathcal{D}})^{T}I_{1}(\beta^{\mathcal{D}}_{0})(\hat{\beta}^{\mathcal{D}}-\tilde{\beta}^{\mathcal{D}})+o_p(1).$$
}
\no{\bf Proof.}\quad
A Taylor's expansion of
$PQ_n(\hat{\beta}^{\mathcal{D}}|V)-PQ_n(\tilde{\beta}^{\mathcal{D}}|V)$ at the
point $\hat{\beta}^{\mathcal{D}}$ yields
$$PQ_n(\hat{\beta}^{\mathcal{D}}|V)-PQ_n(\tilde{\beta}^{\mathcal{D}}|V)=T_1+T_2+T_3+T_4,$$
where \begin{eqnarray*}
        T_1 &=& \nabla PQ_n(\hat{\beta}^{\mathcal{D}}|V)(\hat{\beta}^{\mathcal{D}}-\tilde{\beta}^{\mathcal{D}}), \\
        T_2 &=& -\frac{1}{2}(\hat{\beta}^{\mathcal{D}}-\tilde{\beta}^{\mathcal{D}})^{T}\nabla^2L_n(\hat{\beta}^{\mathcal{D}})(\hat{\beta}^{\mathcal{D}}-\tilde{\beta}^{\mathcal{D}}), \\
        T_3 &=& \frac{1}{6}\nabla^{T}\{(\hat{\beta}^{\mathcal{D}}-\tilde{\beta}^{\mathcal{D}})^{T}\nabla^2L_n(\beta^{\mathcal{D}*})(\hat{\beta}^{\mathcal{D}}-\tilde{\beta}^{\mathcal{D}})\}(\hat{\beta}^{\mathcal{D}}-\tilde{\beta}^{\mathcal{D}}), \\
        T_4
        &=&-\frac{1}{2}(\hat{\beta}^{\mathcal{D}}-\tilde{\beta}^{\mathcal{D}})^{T}\nabla^2P_{\lambda}(\hat{\beta}^{\mathcal{D}})\{I+o(I)\}(\hat{\beta}^{\mathcal{D}}-\tilde{\beta}^{\mathcal{D}}),
      \end{eqnarray*}
where
$P_{\lambda}(\beta)=n\sum_{j=1}^{p}\tilde{p}_{\lambda}(|\beta_{j}|)$.
Since $T_1=0$ as $\nabla PQ_n(\hat{\beta}^{\mathcal{D}}|V)=0$.  By
Lemma 3,  it holds
\begin{equation}\label{condition11}
    \hat{\beta}^{\mathcal{D}}-\tilde{\beta}^{\mathcal{D}}=\Theta^{-1/2}_n\{I_{d+s}-\Theta^{1/2}_nD(D^{T}\Theta_nD)^{-1}D^{T}\Theta^{1/2}_n\}\Theta^{-1/2}_n\Phi_n+o_p(n^{-1/2}),
\end{equation}
where $\Theta_n=I_1(\beta^{\mathcal{D}}_{0})$ and $\Phi_n=\frac{1}{n}\nabla
L_n(\beta^{\mathcal{D}}_{0})$. Note that
$I_{d+s}-\Theta^{1/2}_nD(D^{T}\Theta_nD)^{-1}D^{T}\Theta^{1/2}_n$
is an idempotent matrix with rank $d$. Hence, by a standard argument
and condition (B), we have
$\|\hat{\beta}^{\mathcal{D}}-\tilde{\beta}^{\mathcal{D}}\|=O_p(\sqrt{\frac{d}{n}})$.
Thus, by condition (C), we have
\begin{eqnarray*}
  |T_3| &=& \frac{1}{6}|\sum_{j,k,l}\frac{\partial^3 L_n(\beta^{\mathcal{D}*})}{\partial \beta_{j}\partial \beta_{k}\partial \beta_{l}}(\hat{\beta}^{\mathcal{D}}-\tilde{\beta}^{\mathcal{D}})_j(\hat{\beta}^{\mathcal{D}}-\tilde{\beta}^{\mathcal{D}})_k(\hat{\beta}^{\mathcal{D}}-\tilde{\beta}^{\mathcal{D}})_l| \\
   &\leq&\frac{1}{6}\{\sum_{j,k,l}\sum_{i=1}^n
   M^2_{jkl}(V_i)\}^{1/2}\sqrt{n}\|\hat{\beta}^{\mathcal{D}}-\tilde{\beta}^{\mathcal{D}}\|^3\\
   &=&O_p((np^3)^{1/2})\sqrt{n}O_p((d/n)^{3/2})=O_p(\sqrt{n}p^{3/2}d^{3/2})=o_p(1).
\end{eqnarray*}
Again by condition $b_n=o_p(1/\sqrt{p})$, we have $\|T_4\|^2\leq
nb_n\|\hat{\beta}^{\mathcal{D}}-\tilde{\beta}^{\mathcal{D}}\|^2=no_p(1/\sqrt{p})O_p(d/n)=o_p(1)$.
Thus,
$PQ_n(\hat{\beta}^{\mathcal{D}}|V)-PQ_n(\tilde{\beta}^{\mathcal{D}}|V)=T_2+o_p(1)$.
By condition (B), it is easy to see that
$\|\frac{1}{n}\nabla^2L_n(\hat{\beta}^{\mathcal{D}})+I_1(\beta^{\mathcal{D}}_0)\|=o_p(1/\sqrt{p})$.
Hence, we have $
    \frac{1}{2}(\hat{\beta}^{\mathcal{D}}-\tilde{\beta}^{\mathcal{D}})^{T}\{\nabla^2L_n(\hat{\beta}^{\mathcal{D}})+nI_{1}(\beta_{0}^{\mathcal{D}})\}(\hat{\beta}^{\mathcal{D}}-\tilde{\beta}^{\mathcal{D}})\leq
    o_{p}(n\frac{1}{\sqrt{p}})O_p(d/n)=o_p(1)$.
Thus,
$PQ_n(\hat{\beta}^{\mathcal{D}}|V)-PQ_n(\tilde{\beta}^{\mathcal{D}}|V)=\frac{n}{2}(\hat{\beta}^{\mathcal{D}}-\tilde{\beta}^{\mathcal{D}})^{T}I_{1}(\beta^{\mathcal{D}}_{0})(\hat{\beta}^{\mathcal{D}}-\tilde{\beta}^{\mathcal{D}})+o_p(1)$.

\no\textbf{Proof of Theorem~\ref{theorem1}.}

\no{\bf Proof.}\quad 
Substituting (\ref{condition11}) into Lemma
4, we obtain
\begin{equation}\label{eq4}
  PQ_n(\hat{\beta}^{\mathcal{D}}|V)-PQ_n(\tilde{\beta}^{\mathcal{D}}|V)=\frac{n}{2}\Phi^{T}_n\Theta^{-1/2}_n\{I_{d+s}-\Theta^{1/2}_nD(D^{T}\Theta_nD)^{-1}D^{T}\Theta^{1/2}_n\}\Theta^{-1/2}_n\Phi_n+o_p(1).
\end{equation}
Since
$I_{d+s}-\Theta^{1/2}_nD(D^{T}\Theta_nD)^{-1}D^{T}\Theta^{1/2}_n$
is an idempotent matrix with rank $d$, we can rewritten it as the
product form $A^{T}_nA_n$, where $A_n$ is a $d\times (d+s)$ matrix
that satisfies $A_nA^{T}_n=I_d$. As in the proof of Lemma 2, we can show that
$\sqrt{n}A_n\Theta^{-1/2}_n\Phi_n \rightarrow N(0, I_{d})$. Thus,
\begin{eqnarray*}
  T_n &=& 2\{\sup_{\Omega}PQ_n(\beta|V)-\sup_{\Omega, \beta_{1}=0}PQ_n(\beta|V)\}= 2\{PQ_n(\hat{\beta}^{\mathcal{D}}|V)-PQ_n(\tilde{\beta}^{\mathcal{D}}|V)\}\\
   &=&
   (\sqrt{n}\Theta^{-1/2}_n\Phi_n)^{T}(I_{d+s}-\Theta^{1/2}_nD(D^{T}\Theta_nD)^{-1}D^{T}\Theta^{1/2}_n)(\sqrt{n}\Theta^{-1/2}_n\Phi_n)+o_p(1)\\
   &=&\{\sqrt{n}A_n\Theta^{-1/2}_n\Phi_n\}^{T}\{\sqrt{n}A_n\Theta^{-1/2}_n\Phi_n\}+o_p(1) \rightarrow \chi^2_d.
\end{eqnarray*}

\no\textbf{Proof of Theorem~\ref{theorem2}.}

\no{\bf Proof.}\quad Let $\tilde{\delta}=(\delta^{T},0_{s})^{T}$ be a $(d+s)\times 1$
vectors. If $H_1: \beta_{01}=\delta n^{-1/2}$ is true,
denote
$\tilde{\beta}^{\mathcal{D}}_n=(\beta^{T}_{01},\beta^{T}_{021})^{T}$ be a
sequence of ''true" values of $\beta^{\mathcal{D}}$. Consider
Taylor series expansion of $\nabla L_n(\beta^{\mathcal{D}}_{0})$ about
$\beta^{\mathcal{D}}=\tilde{\beta}^{\mathcal{D}}_n$,
$$\nabla
L_n(\beta^{\mathcal{D}}_{0})=\nabla L_n(\tilde{\beta}^{\mathcal{D}}_n)+
\nabla^2 L_n(\tilde{\beta}_*^{\mathcal{D}})(\beta^{\mathcal{D}}_{0}-\tilde{\beta}^{\mathcal{D}}_n)=\nabla L_n(\tilde{\beta}^{\mathcal{D}}_n)+n^{1/2}(I_1(\beta^{\mathcal{D}}_{0})+o_p(1))\tilde{\delta},$$
where $\tilde{\beta}_*^{\mathcal{D}}$ such that
$\|\tilde{\beta}_*^{\mathcal{D}}-\beta^{\mathcal{D}}_{0}\|<\|\beta^{\mathcal{D}}_{0}-\tilde{\beta}^{\mathcal{D}}_n\|$.
Continue the notation of Lemma 4, and by the proof of Lemma 2, we
have $n^{-1/2}A_n\Theta^{-1/2}_n\nabla L_n(\tilde{\beta}^{\mathcal{D}}_n)
\rightarrow N_d(0,G)$, where $A_n$ is a $d\times (d+s)$ matrix
such that $A_nA^{T}_n\rightarrow G$, and $G$ is a $d\times d$
nonnegative symmetric matrix. Thus,
\begin{equation}\label{aeq5}
\sqrt{n}A_n\Theta^{-1/2}_n\Phi_n=n^{-1/2}A_n\Theta^{-1/2}_n\nabla
L_n(\beta_0^{\mathcal{D}}) \rightarrow
N_d(A_n\Theta^{1/2}_n\tilde{\delta},G).
\end{equation} Thus, let $A_n$ be the
$d\times (d+s)$ matrix such that
$A^{T}_nA_n=I_{d+s}-\Theta^{1/2}_nD(D^{T}\Theta_nD)^{-1}D^{T}\Theta^{1/2}_n$,
and $A_nA^{T}_n=I_d$, it holds that $\sqrt{n}A_n\Theta^{-1/2}_n\Phi_n
\rightarrow N_{d}(A_n\Theta^{1/2}_n\tilde{\delta}, I_d)$. And finally,
\begin{eqnarray*}
  T_n &=& 2\{\sup_{\Omega}PQ_n(\beta|V)-\sup_{\Omega, \beta_{1}=0}PQ_n(\beta|V)\}=2\{PQ_n(\hat{\beta}^{\mathcal{D}}|V)-PQ_n(\tilde{\beta}^{\mathcal{D}}|V)\}\\
   &=&
   (\sqrt{n}\Theta^{-1/2}_n\Phi_n)^{T}(I_{d+s}-\Theta^{1/2}_nD(D^{T}\Theta_nD)^{-1}D^{T}\Theta^{1/2}_n)(\sqrt{n}\Theta^{-1/2}_n\Phi_n)+o_p(1)\\
   &=&\{\sqrt{n}A_n\Theta^{-1/2}_n\Phi_n\}^{T}\{\sqrt{n}A_n\Theta^{-1/2}_n\Phi_n\}+o_p(1) \rightarrow
   \chi^2_d(\gamma),
\end{eqnarray*}
where $\gamma=\tilde{\delta}\Theta^{1/2}_n
A^{T}_nA_n\Theta^{1/2}_n\tilde{\delta}=\delta^{T}C_{11.2}\delta$,
with $C_{11.2}$ is defined in Theorem \ref{theorem2}.

\bibliographystyle{abbrvnat}
\bibliography{ex}
\end{document}